\documentclass[english]{elsarticle}
\usepackage{mathptmx}
\usepackage[LGR,T1]{fontenc}
\usepackage[latin9]{inputenc}
\usepackage{geometry}
\usepackage{amsmath}
\usepackage{amssymb}
\usepackage{graphicx}
\usepackage{esint}

\makeatletter

\DeclareRobustCommand{\greektext}{%
  \fontencoding{LGR}\selectfont\def\encodingdefault{LGR}}
\DeclareRobustCommand{\textgreek}[1]{\leavevmode{%
  \IfFileExists{grtm10.tfm}{}{\fontfamily{cmr}}\greektext #1}}
\DeclareFontEncoding{LGR}{}{}
\DeclareTextSymbol{\~}{LGR}{126}

\date{}

\makeatother

\usepackage{babel}
\begin{document}

\title{Genuinely Multidimensional Kinetic Scheme For Euler Equations}

\author{Praveer Tiwari$^{1}$, SV Raghurama Rao$^{2}$}

\address{1. Department of Physics, Indian Institute of Science, Bangalore }

\address{2. Department of Aerospace Engineering, Indian Institute of Science,
Bangalore}
\begin{abstract}
A new framework based on Boltzmann equation which is genuinely multidimensional
and mesh-less is developed for solving Euler's equations. The idea
is to use the method of moment of Boltzmann equation to operate in
multidimensions using polar coordinates. The aim is to develop a framework
which is genuinely multidimensional and can be implemented with different
methodologies, no matter whether it is in finite difference, finite
volume or finite element form. There is a considerable improvement
in capturing shocks and other discontinuities. Also, since the method
is multidimensional, the flow features are captured isotropically.
The method is further extended to second order using 'Arc of Approach'
concept. The framework is developed as a finite difference method
(called as GINEUS) and is tested on the benchmark test cases. The
results are compared against Kinetic Flux Vector Splitting Method.
\end{abstract}
\maketitle

\section{Introduction}

With the emergence of Gudonov's scheme (\citet{Godunov}), the quest
for finding scheme to resolve shocks and other discontinuities for
hyperbolic gas dynamic equations had started. What followed are several
central and upwind discretization methods, of which, schemes by \citet{Leer1979},
\citet{Osher1982} and \citet*{Wendroff1960} are some of the prominent
ones. A feat of exact shock capturing in 1D is achieved in the schemes
by \citet{Roe1981}, \citet{Liou1996}(AUSM+) and \citet{Jaisankar2009}(MOVERS)
of which AUSM+ also achieves exact contact discontinuity capturing.
Although, these schemes were efficient, robust and accurate in 1D,
their 2D counterparts were not accurate in resolving flow features
because they were modelling waves propagating along the grid. Physically,
the waves can propagate in any direction, so splitting it along grid
axes becomes meaningless. Hence, it was realized that there is a need
of a new framework which could account for the wave propagation in
multidimensions. This led to development of a new class of schemes
which are genuinely multidimensional that can resolve shocks and other
discontinuities even when they are not aligned to grids.

Several attempts have been made to eliminate the grid dependence for
capturing flow features. \citet{P.L.Roe1982} laid the foundation
of this class of schemes by introducing fluctuation splitting method
which was based on the idea that the solution assumes piecewise continuous
solution inside a cell, contrary to Gudonov's framework. This work
was further developed to propose a method (\citet{Roe1986}) where
the multidimensional Euler equations are modified, in a similar way
as it was done for the Riemann solver (\citet{Roe1981}), to simpler
components. \citet{CHHirsch1987} introduced a scheme in which they
diagonalized the Euler equations using the local wave propagation
direction coming from pressure gradient and stress tensor. This way
upwinding is done based on local flow features which is independent
of the background grid. Building on this idea, \citet{Powell1989a}
introduced genuinely multidimensional cell-vertex scheme in which
data is decomposed into four variables which is then convected to
appropriate directions using cell vertex scheme. \citet{IjazParpia1990}
introduced a scheme based on the assumption of three waves at each
interface where the direction and strength of these waves are determined
by local flow features. \citet{Colella1990} introduced a multidimensional
scheme where numerical fluxes are evaluated based on characteristic
form of multidimensional Euler equation at the interface. \citet{Dadone1991}
introduced a scheme based on flux-difference splitting, at each face,
along two perpendicular direction where the directions are chosen
based on pressure gradient at the face. Roe came back with another
idea (\citet{Roe1991}) where linear wave solutions are obtained from
the piecewise linear data over the cell (which is the enclosure of
the data points) and hence avoiding the need of flow variables at
the cell interfaces. \citet{IjazParpia1992}, introduced yet another
scheme which extracts the information about the waves using reconstruction
procedure applied on the data present at the vertices of a triangle.
\citet{Abgrall1993} came up with a scheme invloving linearization
of Euler equation whose exact solution is then used for evaluating
the flux at the interfaces in multidimensional way.\citet{HDeconinck1993}
deviced a multidimensional scheme based on fluctuation-splitting scheme
introduced by \citet{P.L.Roe1982}. Later that year, \citeauthor{W.Eppard1993}
took Deconinck's idea to Boltzmann level. They used fluctuation splitting
of distribution function at kinetic level thereby obtaining the fluctuations
in conserved variables as its moments. \citet{LeVeque1997} introduced
a class of high resolution multidimensional schemes based on solving
Riemann problems and using limiters to get the resulting waves which
is then propagated in multidimensions. \citet{AlexanderKurganov2001}
introduced yet another genuinely multidimensional scheme using precise
information about the local speeds of propagation in his high resolution
central scheme \citet{A.Kurganov2000}. \citet{SVRaghuramaRao2003}
introduced framework which involved using relaxation schemes combined
with splitting method to linearize the non linear hyperbolic conservation
equation followed by tracing the foot of the characteristic of the
resulting equation to get the genuinely multidimensional structure.
\citet{R.Kissmann2009} extended this idea to arbitrary orthogonal
grids. Lucacova coupled finite volume formulation with operators constructed
using bicharacteristics of the multidimensional hyperbolic system
constrained to capture the wave propagation in arbitrary direction
( \citet{Lukacova-Medvidova2002} and \citet{Lukacova-Medvidova2004}).
Later, \citet{K.R.Arun2009} extended this idea to hyperbolic systems
for spatially varying flux functions. \citet{Rossmanith2008} extended
Leveque and Pelanti's idea (\citet{LeVeque2001}) to construct a multidimensional
relaxation system to obtain multidimensional approximate Riemann solvers.
\citet{S.E.Razavi2008} extracted the multidimensional characteristic
structure of incompressible flows, modified by artificial compressibility,
to construct inherent multidimensional upwind scheme. \citet{Balsara2010}
introduced a multidimensional Riemann solver that takes input from
neighbouring points for getting fluxes in a way which creates self-similar
strongly interacting one dimensional Riemann problem. Initially, the
scheme was restricted to use only HLLE Riemann solver, but later he
himself developed it to work with any 1D self-similar Riemann solver
in structured and unstructured mesh (\citet{Balsara2014} and \citet{DinshawS.Balsara2015}
respectively). Next on the list comes \citet{Mishra2011}, who reformulated
finite volume scheme in terms of vertex centered numerical potentials
to get a genuinely multidimensional structure. One of the recent is
by K.R Arun and M. Lukacova-Medvid'ova (\citealp{K.R.Arun2013}) where
they work in DVBE framework and isotropically cover the domain by
foot of the characteristic. \citet{S.Jaisankar2013} used direction
based diffusion regulator with dimension splitting solvers to moderate
the excess multidimensional diffusion.

Instead of introducing a new framework to model multidimensionality
of waves, the innate nature of method of moments of Boltzmann equation
can be exploited to capture waves in multidimensions. This feature
has been explored by \citet{SVRaghuramaRao1991} where he used rotational
coordinates to take moments of the Boltzmann equation while coupling
the upwinding with the direction of microscopic flow velocity. Although,
this was a good attempt to extract the genuinely multidimension nature
of kinetic framework, the absence of closed form expressions for the
integral meant that scheme would have to depend upon numerical quadrature
which was costly. The present work aims to re-introduce the idea in
a way which is computationally inexpensive and technically more obvious.
The aim of the paper is to introduce the idea in upwind framework.

Next section explains briefly, the method of moments of Boltzmann
equation to obtain Euler equations. The following sections will introduce
the first order formulation of the idea, the second order extension
and ultimately the last section will give the numerical results for
the standard test cases.

\section{Euler Equations from Boltzmann Equation}

The Boltzmann Equation, given by ($\ref{eq:BE}$), has proved to be
useful tool in analysis of macroscopic fluid flows. 

\begin{equation}
\frac{\partial f}{\partial t}+\mathbf{v}\cdot\mathbf{\overrightarrow{\bigtriangledown}}_{x}f+\mathbf{a}\cdot\mathbf{\mathbf{\overrightarrow{\bigtriangledown}}}_{v}f=J(\mathbf{x},\mathbf{v},t)\label{eq:BE}
\end{equation}

where $f$ is the probability distribution function, $\mathbf{v}=(v_{x,}v_{y},v_{z})$,
$\mathbf{a}$ is the acceleration due to external forces and $J$
is the collision term. In this section, we will briefly discuss about
obtaining Euler Equation (inviscid form of N-S E) from Boltzmann Equation.
Consider equation (1) and the vector 

\begin{equation}
\psi=\begin{Bmatrix}1\\
v_{x}\\
v_{y}\\
v_{z}\\
I+\frac{1}{2}(v_{x}^{2}+v_{y}^{2}+v_{z}^{2})
\end{Bmatrix}.
\end{equation}

Taking moment of (1) with $\psi$, we have

\begin{equation}
<\psi,\frac{\partial f}{\partial t}+\mathbf{v}\cdot\mathbf{\overrightarrow{\bigtriangledown}}_{x}f+\mathbf{a}\cdot\mathbf{\mathbf{\overrightarrow{\bigtriangledown}}}_{v}f=J(\mathbf{x},\mathbf{v},t)>
\end{equation}
 It is assumed that the collisions between fluid molecules at microscopic
level are perfectly elastic so that the collision integral conserves
mass, momentum and energy. Secondly, it is assumed that there is no
external forces acting on the system which removes the third term
on the right hand side of equation $\ref{eq:BE}$. In this setup,
system relaxes to Maxwellian distribution, given by

\begin{equation}
f^{M}=\rho\left(\frac{\beta}{\pi I_{0}}\right)^{\frac{D}{2}}e^{-\beta\sum^{D}(v_{i}-u_{i})^{2}-\frac{I}{I_{0}}},
\end{equation}

where $\beta=1/2RT$ , $D$ is the spatial dimension of the fluid
flow and $I$ is the internal energy.

Accounting for these two assumptions, the moment equation reads:

\begin{equation}
<\psi,\frac{\partial f}{\partial t}+\mathbf{v}\cdot\mathbf{\overrightarrow{\bigtriangledown}}_{x}f=0,\!\, f=f^{M}>.\label{eq:Moment}
\end{equation}

The above equation boils down to

\begin{equation}
\frac{\partial U}{\partial t}+\frac{\partial G_{1}}{\partial x}+\frac{\partial G_{2}}{\partial y}+\frac{\partial G_{3}}{\partial z}=0,\label{eq:Euler3D}
\end{equation}

where 

\begin{equation}
U=\begin{Bmatrix}\rho\\
\rho u_{x}\\
\rho u_{y}\\
\rho u_{z}\\
\rho E
\end{Bmatrix},G_{1}=\begin{Bmatrix}\rho u_{x}\\
P+\rho u_{x}^{2}\\
\rho u_{x}u_{y}\\
\rho u_{x}u_{z}\\
\rho u_{x}E
\end{Bmatrix},G_{2}=\begin{Bmatrix}\rho u_{y}\\
\rho u_{x}u_{y}\\
P+\rho u_{y}^{2}\\
\rho u_{y}u_{z}\\
\rho u_{y}E
\end{Bmatrix},G_{3}=\begin{Bmatrix}\rho u_{z}\\
\rho u_{x}u_{z}\\
\rho u_{y}u_{z}\\
P+\rho u_{y}^{2}\\
\rho u_{z}E
\end{Bmatrix},
\end{equation}

$E=e+\sum_{i=x,y,z}(u_{i}^{2}/2)$ and $e=P/\rho(\gamma-1)$.

The hyperbolic system $(\ref{eq:Euler3D})$ represents the well-known
Euler equations of Gas Dynamics. In 2D, it boils down to

\begin{equation}
\frac{\partial U}{\partial t}+\frac{\partial G_{1}}{\partial x}+\frac{\partial G_{2}}{\partial y}=0,\label{eq:Euler2D}
\end{equation}

where 

\begin{equation}
U=\begin{Bmatrix}\rho\\
\rho u_{x}\\
\rho u_{y}\\
\rho E
\end{Bmatrix},\, G_{1}=\begin{Bmatrix}\rho u_{x}\\
P+\rho u_{x}^{2}\\
\rho u_{x}u_{y}\\
\rho u_{x}E
\end{Bmatrix},\, G_{2}=\begin{Bmatrix}\rho u_{y}\\
\rho u_{x}u_{y}\\
P+\rho u_{y}^{2}\\
\rho u_{y}E
\end{Bmatrix}.
\end{equation}

\section{Grid Independent Essentially Upwinding Scheme}

\begin{figure}
\begin{centering}
\includegraphics[scale=0.2]{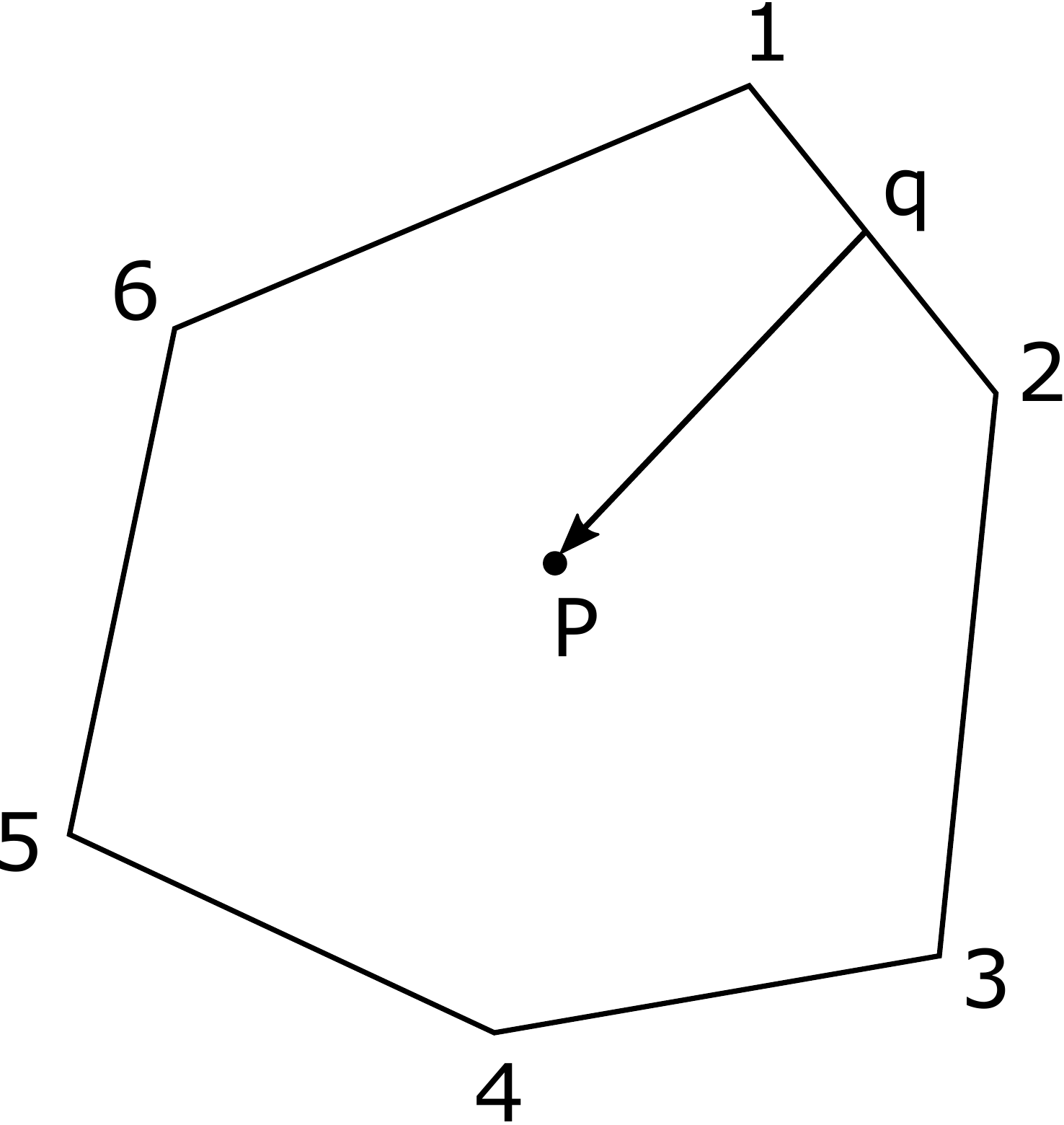}
\par\end{centering}

\caption{Central point P with all the neigbouring points\label{fig:Central-point}.}
\end{figure}

The present scheme basically does upwinding at microscopic level based
on particle velocity. It is an ingenious extension, in some sense,
of the grid-aligned kinetic theory based upwind method (KFVS). We
have a central point P whose value has to be updated in time. The
neighbours around the central point P are provided (as shown in the
figure ($\ref{fig:Central-point}$)).

Consider the face RS such that R and S are in the set of data points
of the neighbourhood of P.

\begin{figure}[h]
\begin{centering}
\includegraphics[scale=0.3]{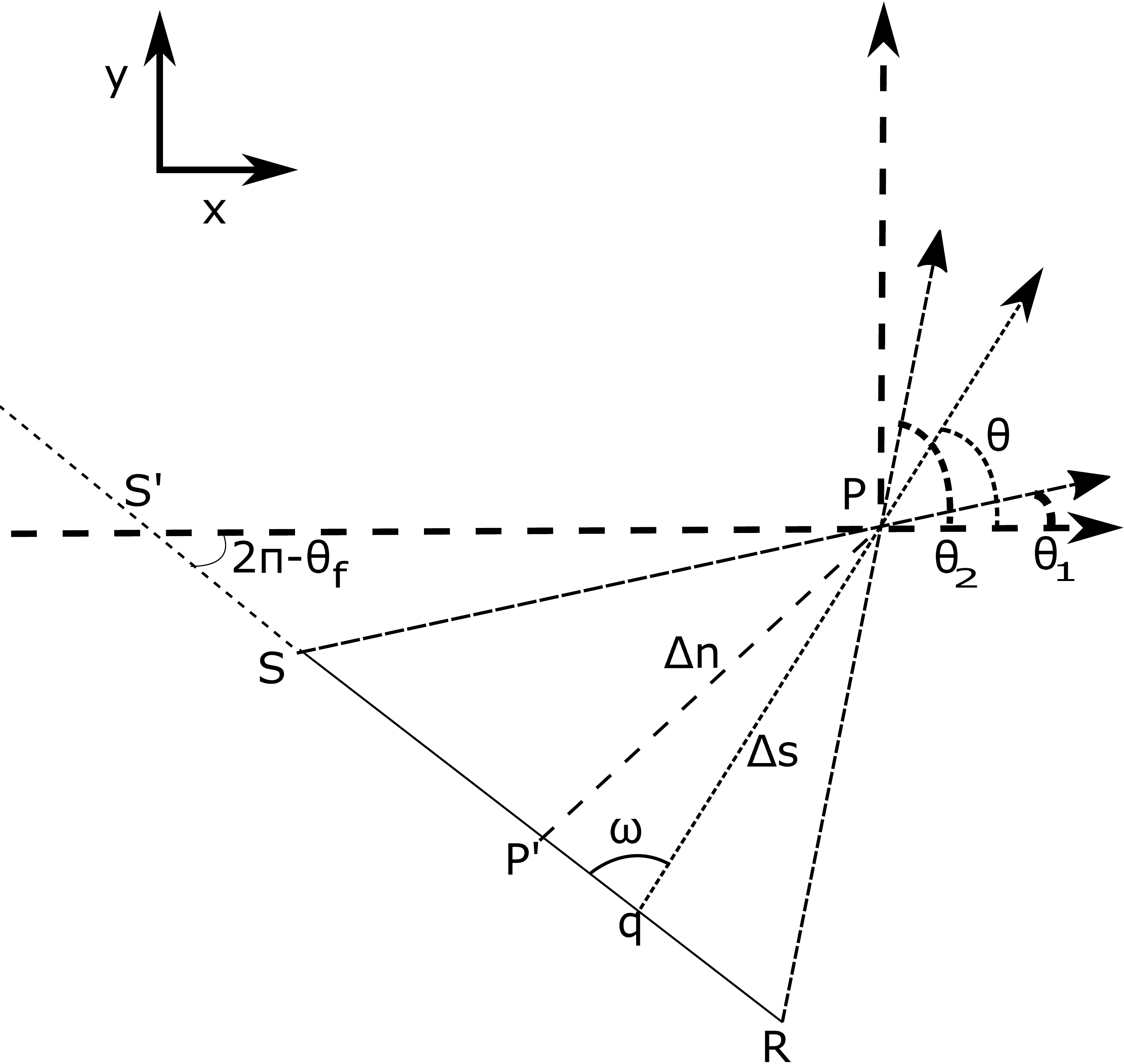}
\par\end{centering}

\caption{Upwinding in Multidimensions.}
\end{figure}

The flux term of the Boltzmann equation is given by :

\begin{equation}
\overrightarrow{v}.\nabla_{x}f=v\frac{\partial f}{\partial s},\label{eq:Flux}
\end{equation}

where $\hat{s}$ is the direction of velocity $\overrightarrow{v}$.
The above equation can be proved in two steps. First note that dot
product is a scalar and hence is invariant under rotation of coordinate
axes. In the second step, choose the coordinate axes such that one
of the axis is parallel to $\overrightarrow{v}$.

\subsection{First Order Formulation}

Equation $(\ref{eq:Flux})$ can be approximated as following: 

\begin{equation}
v\frac{\partial f}{\partial s}\backsimeq v\frac{\bigtriangleup f}{\bigtriangleup s}
\end{equation}

or

\begin{equation}
v\frac{\bigtriangleup f}{\bigtriangleup s}=v\frac{\bigtriangleup f}{\bigtriangleup n}|sin\omega|
\end{equation}

Since polar coordinates is being used, the moment vector is written
as:

\begin{equation}
\psi=\begin{Bmatrix}1\\
vcos\theta\\
vsin\theta\\
I+\frac{1}{2}v^{2}
\end{Bmatrix}
\end{equation}

Using equation $(\ref{eq:Moment})$, moment equation reads

\begin{equation}
<\psi,v\frac{\partial f}{\partial s},\! f=f^{M}>\backsimeq\intop_{0}^{\infty}\intop_{0}^{2\pi}\psi v^{2}\left[\frac{f_{p}^{M}-f_{q}^{M}}{\bigtriangleup n}\right]sin\omega\, dvd\theta\label{eq:Moment_Int}
\end{equation}

Note that we have already integrated out internal energy because it
has very little role to play in upwinding.

The next step is to construct a polygon with neighboring points as
its vertices. Then, the above integral can be written as the summation
of integrals along different edges. Equation $(\ref{eq:Moment_Int})$,
then, reads as:

\begin{equation}
<\psi,v\frac{\partial f}{\partial s},\! f=f^{M}>\,\,\,\backsimeq\sum_{all\, edges}\intop_{0}^{\infty}\intop_{\theta_{1,edge}}^{\theta_{2,edge}}\psi v^{2}\frac{f_{p}^{M}-f_{q}^{M}}{\bigtriangleup n}sin\omega\, dvd\theta\label{eq:Moment_Int_1}
\end{equation}

where $\theta_{1,edge}$ and $\theta_{2,edge}$ are the starting and
ending angle (going anticlockwise) of the edge, when measured with
center point as origin. The expression for $f_{q}^{M}$ as a function
of the polar angle can be obtained by doing linear interpolation of
the distribution function along the edge under consideration. So $f_{q}^{M}$
can be written as:

\begin{equation}
f_{q}^{M}=D_{1}(\theta)f_{R}^{M}+D_{2}(\theta)f_{S}^{M}
\end{equation}

where $R$ and $S$ are the vertices corresponding to the given edge.
The expressions for $D_{1}(\theta)$ and $D_{2}(\theta)$ are derived
in Appendix. Finally, note that the integral given by $(\ref{eq:Moment_Int_1})$
has no closed form solution for Maxwellian distribution of velocity
at equilibrium. This means that, to solve $(\ref{eq:Moment_Int_1})$,
numerical quadrature has to be used. This problem can be avoided by
noting the work of Sanders and Prendergast \citep{R.H.Sanders1974},
in which he used combinations of Dirac-delta function at different
points. In other words, Maxwellian can be approximated by 5 beams
(for 2-D and 7 beams in 3-D). More precisely it is written as

\begin{equation}
f^{M}\simeq e^{-\frac{I}{I_{0}}}(a\delta(v_{1}-u_{1})\delta(v_{2}-u_{2})+b\sum_{(u^{(1)},u^{(2)})\,\varepsilon\,\text{\textgreek{W}}}(\delta(v_{1}-u^{(1)})\delta(v_{2}-u^{(2)}))
\end{equation}
where $\Omega$= \{($u_{1},\, u_{2}$), ($u_{1},\, u_{2}$$\pm\bigtriangleup u$),
($u_{1}\pm\bigtriangleup u,\, u_{2}$)\}. Call this approximate function
$f^{B}$. The values of a, b and $\bigtriangleup u$ are derived by
ensuring that it gives correct moments (Appendix).

Using this, the semi-discrete equation is given by:

\begin{equation}
\frac{\partial U}{\partial t}=\sum_{all\, edges}\frac{1}{\bigtriangleup n_{RS}}\intop_{0}^{\infty}\intop_{\theta_{1,RS}}^{\theta_{2,RS}}\psi v^{2}sin\alpha(D_{1}(\theta)f_{R}^{B}+D_{2}(\theta)f_{S}^{B}-f_{p}^{B})dvd\theta
\end{equation}

The above integral for each edge of the sum contains 15 integrals
(3 distribution functions with 5 beams each) of the form

\begin{equation}
\intop_{0}^{\infty}\intop_{\theta_{1,RS}}^{\theta_{2,RS}}\frac{\psi v^{2}\times D(\theta)\times c\times sin\alpha\times\delta(vcos\theta-u^{(1)})\delta(vsin\theta-u^{(2)})}{\bigtriangleup n_{RS}}dvd\theta,\label{eq:Final_Int}
\end{equation}

where $D(\theta)=-1$ for $f_{p}^{B}$ ; $D_{1}(\theta)$ for $f_{R}^{B}$
and $D_{2}(\theta)$ for $f_{S}^{B}$. Similarly, $c=a$ or $b$,
depending on whether it is a central beam or a side beam. And ($u^{(1)},\, u^{(2)}$)
= ($u_{1},\, u_{2}$) or ($u_{1},\, u_{2}$$\pm\bigtriangleup u$)
or ($u_{1}\pm\bigtriangleup u,\, u_{2}$) depending on the beam being
evaluated.

Integral $(\ref{eq:Final_Int})$ comes out to be

\begin{equation}
I=\begin{cases}
\frac{\phi qD(\theta_{0})csin\omega}{\bigtriangleup n_{RS}} & \theta_{1,RS}<\theta_{0}<\theta_{2,RS}\\
0 & otherwise
\end{cases},
\end{equation}

where $q=\sqrt{(u^{(1)})^{2}+(u^{(2)})^{2}}$, $\theta_{0}$ is the
angle made ($u^{(1)},\, u^{(2)}$) with positive x-axis and 

\begin{equation}
\phi=\begin{Bmatrix}1\\
u^{(1)}\\
u^{(2)}\\
I+\frac{1}{2}q^{2}
\end{Bmatrix}.
\end{equation}

\subsection{Second Order Formulation}

In this section the formulation given in section 3.1 is extended to
second order. From$(\ref{eq:Flux})$, we have

\begin{equation}
\overrightarrow{v}.\nabla_{x}f=v\frac{\partial f}{\partial s}
\end{equation}

For doing second order approximation of $\frac{\partial f}{\partial s}$
, 5 point stencil is required in each direction. The concept ``Arc
of Approach'' is introduced for getting the same. The idea is to
use the second order discretization of linear convection equation
and apply it to the Boltzmann equation which has been reduced to one
dimension by equation ($\ref{eq:Flux}$). 

The 1D linear convection equation is given by

\begin{figure}
\begin{centering}
\includegraphics[scale=0.3]{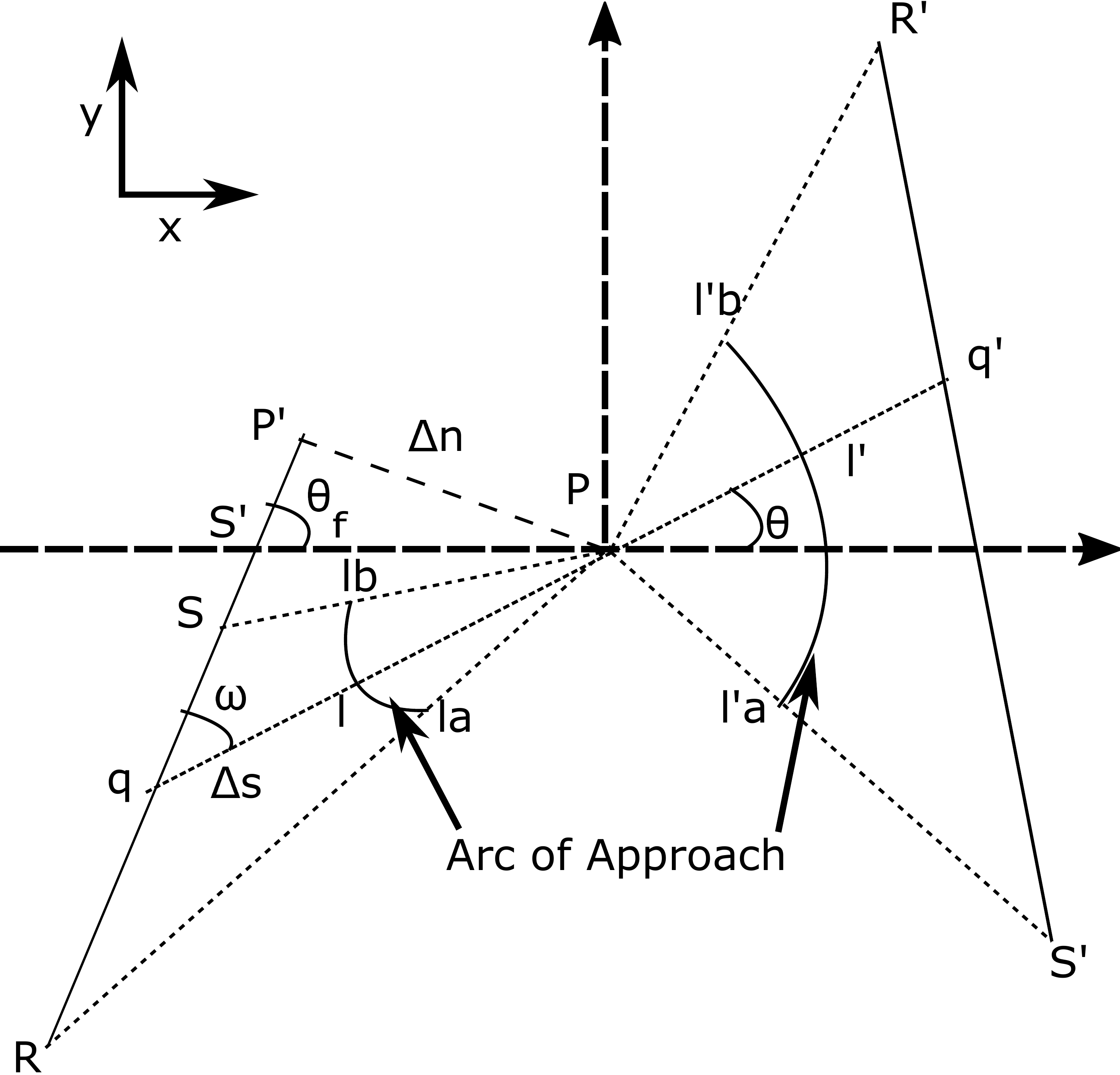}
\par\end{centering}

\caption{Showing different interpolations required for the Arc of Approach
method. }
\end{figure}

\begin{equation}
u_{t}+au_{x}=0\,\,\,\,\, a>0.
\end{equation}

The update formula according to common framework for looking at limiters
given by \citet{Sweby1984}, reads

\begin{equation}
u_{i}^{n+1}=u_{i}^{n}-\lambda(u_{i}-u_{i-1})-(\psi_{i}F_{i+1/2}-\psi_{i-1}F_{i-1/2})\label{eq:Sweby_update}
\end{equation}

where $\lambda=\frac{a\bigtriangleup t}{\bigtriangleup x}$,

\begin{equation}
F_{i+1/2}=\frac{1}{2}\lambda(1-\lambda)(u_{i+1}-u_{i})
\end{equation}

and
\begin{equation}
\psi_{i}=\psi(R_{i})=\psi\left(\frac{u_{i+1}-u_{i}}{u_{i}-u_{i-1}}\right)
\end{equation}

is the limiter function.

Since Boltzmann equation is also being discretized in one direction,
its update formula for the second order extension can be written in
the form of $\ref{eq:Sweby_update}$ . The central point P assumes
the role of $i$ and all other points are assigned along the line
segment $qPq'$. 

The fully discretized update formula (following equation $(\ref{eq:Sweby_update})$),
is then given by

\begin{equation}
f_{p}^{n+1}=f_{P}^{n}-\frac{1}{2}\left[\frac{2v\bigtriangleup t}{\bigtriangleup s}\left(f_{p}-f_{l}\right)+F_{l'P}-F_{Pl}\right],\label{eq:BD}
\end{equation}

where

\begin{equation}
F_{l'P}=\psi_{P}\left(\frac{v\bigtriangleup t}{\bigtriangleup s'}\right)\left(1-\frac{v\bigtriangleup t}{\bigtriangleup s'}\right)\left(f_{l'}-f_{P}\right)
\end{equation}

and

\begin{equation}
F_{Pl}=\psi_{l}\left(\frac{v\bigtriangleup t}{\bigtriangleup s}\right)\left(1-\frac{v\bigtriangleup t}{\bigtriangleup s}\right)\left(f_{P}-f_{l}\right),
\end{equation}

where

\begin{equation}
\psi_{P}=\psi(R_{P})=\psi\left(\frac{\rho_{l'}-\rho_{P}}{\rho_{P}-\rho_{l}}\right)
\end{equation}

and

\begin{equation}
\psi_{l}=\psi(R_{l})=\psi\left(\frac{\rho_{P}-\rho_{l}}{\rho_{l}-\rho_{q}}\right).
\end{equation}

Finally, the distribution function at the point $l$ and $l'$ can
be obtained by aforementioned Arc of Approach method. The method consists
of two consecutive linear interpolations. In the first step, linear
interpolation along the line $PR$ and along the line $PS$ will respectively,
give the distribution function at $lb$ and $la$. The corresponding
expressions are:

\begin{equation}
f_{la}(r)=\left(\frac{f_{S}-f_{R}}{PS}\right)r+f_{p},
\end{equation}

\begin{equation}
f_{lb}(r)=\left(\frac{f_{S}-f_{P}}{PR}\right)r+f_{p},
\end{equation}

where $r$ is the distance from the center. For, simplicity, its value
is chosen to be $\bigtriangleup s/2$. In the second step, the linear
interpolation is done along the arc of radius $\bigtriangleup s/2$.
This gives the distribution function at $l$ and $l'$. The expressions
are given by

\begin{equation}
f_{l}=\left(\frac{f_{lb}-f_{la}}{\theta_{R}-\theta_{S}}\right)\theta+\left(\frac{f_{la}\theta_{R}-f_{lb}\theta_{S}}{\theta_{R}-\theta_{S}}\right)
\end{equation}

Similarly, for the other side of line $Pq$, we have

\begin{equation}
f_{l'}=\left(\frac{f_{l'b}-f_{l'a}}{\theta_{R'}-\theta_{S'}}\right)\theta+\left(\frac{f_{l'a}\theta_{R'}-f_{l'b}\theta_{S'}}{\theta_{R'}-\theta_{S'}}\right).
\end{equation}

Now, after taking moments of equation ($\ref{eq:BD}$) with $\psi$,
we get

\begin{equation}
\frac{\partial U}{\partial t}=-\sum_{all\, edges}\intop_{0}^{\infty}\intop_{\theta_{1,RS}}^{\theta_{2,RS}}\frac{\psi v}{2}\left[\frac{2v\bigtriangleup t}{\bigtriangleup s}\left(f_{p}-f_{l}\right)+F_{l'P}-F_{Pl}\right]dvd\theta
\end{equation}

Simplifying the integrand, we have

\begin{equation}
\frac{\partial U}{\partial t}=-\sum_{all\, edges}\intop_{0}^{\infty}\intop_{\theta_{1,RS}}^{\theta_{2,RS}}\psi v\left[\chi_{P}(v,\theta)f_{P}+\chi_{S}(v,\theta)f_{S}+\chi_{R}(v,\theta)f_{R}+\chi_{S'}(v,\theta)f_{S'}+\chi_{R'}(v,\theta)f_{R'}\right]dvd\theta
\end{equation}

where

\begin{equation}
\begin{bmatrix}\chi_{P}(v,\theta)\\
\chi_{S}(v,\theta)\\
\chi_{R}(v,\theta)\\
\chi_{S'}(v,\theta)\\
\chi_{R'}(v,\theta)
\end{bmatrix}=\begin{bmatrix}B-D+\psi_{p}(C'-A')+\psi_{l}(C-A)\\
A\psi_{l}-B\\
D-C\psi_{l}\\
A'\psi_{P}\\
-C'\psi_{P}
\end{bmatrix}
\end{equation}

where

\begin{equation}
\begin{bmatrix}A\\
B\\
C\\
D
\end{bmatrix}=\begin{bmatrix}\frac{v}{2PS}(1-\frac{v\bigtriangleup t}{\bigtriangleup s})\frac{(\theta-\theta_{R})}{\theta_{S}-\theta_{R}}\\
\frac{v}{PS}\frac{(\theta-\theta_{R})}{\theta_{S}-\theta_{R}}\\
\frac{v}{2PR}(1-\frac{v\bigtriangleup t}{\bigtriangleup s})\frac{(\theta-\theta_{S})}{\theta_{S}-\theta_{R}}\\
\frac{v}{PR}\frac{(\theta-\theta_{S})}{\theta_{S}-\theta_{R}}
\end{bmatrix}
\end{equation}

and

\begin{equation}
\begin{bmatrix}A'\\
B'\\
C'\\
D'
\end{bmatrix}=\begin{bmatrix}\frac{v}{2PS'}(1-\frac{v\bigtriangleup t}{\bigtriangleup s'})\frac{(\theta-\theta_{R'})}{\theta_{S'}-\theta_{R'}}\\
\frac{v}{PS'}\frac{(\theta-\theta_{R'})}{\theta_{S'}-\theta_{R'}}\\
\frac{v}{2PR'}(1-\frac{v\bigtriangleup t}{\bigtriangleup s'})\frac{(\theta-\theta_{S'})}{\theta_{S'}-\theta_{R'}}\\
\frac{v}{PR'}\frac{(\theta-\theta_{S'})}{\theta_{S'}-\theta_{R'}}
\end{bmatrix}.
\end{equation}

The above integral for each edge of the sum contains 25 integrals
(5 distribution functions with 5 beams each) of the form

\begin{equation}
I=\intop_{0}^{\infty}\intop_{\theta_{1,rs}}^{\theta_{2,rs}}\psi v\times\chi(v,\theta)\times c\times\delta(vcos\theta-u^{(1)})\delta(vsin\theta-u^{(2)})dvd\theta\label{eq:Final_Int-1}
\end{equation}

where $\chi$=$\chi_{P}$, $\chi_{S}$, $\chi_{R}$, $\chi_{R'}$or
$\chi_{S'}$ and other variables is as defined for equation ($\ref{eq:Moment_Int}$).
Finally, integrating it gives

\begin{equation}
I=\begin{cases}
\phi\chi(q,\theta_{0})c & \theta_{1,rs}<\theta_{0}<\theta_{2,rs}\\
0 & otherwise
\end{cases}.
\end{equation}

\section{Results and Discussion}

The scheme developed in previous section has been applied on the following
test cases. The first test case is solving 2-D Burgers' Equation given
by

\begin{equation}
\frac{\partial u}{\partial t}+\frac{\partial\frac{u^{2}}{2}}{\partial x}+\frac{\partial u}{\partial y}=0.
\end{equation}

And rest of the test case is for 2D Euler's Equations

\subsection{An oblique shock for 2-D Burgers' equation:}

Specification (\citet{Spekreijse1987}):

Domain: (0,1)$\times$(0,1)

Grid: 32$\times$32, 64$\times$64

Boundary-Conditions (Oblique Shock):

\begin{equation}
\begin{Bmatrix}u(0,y)=1.5\,\,\,0<y<1\\
u(1,y)=-0.5\,\,\,0<y<1\\
u(x,0)=1.5-2x\,\,\,0<x<1
\end{Bmatrix}
\end{equation}

\begin{figure}
\begin{centering}
\includegraphics[scale=0.25]{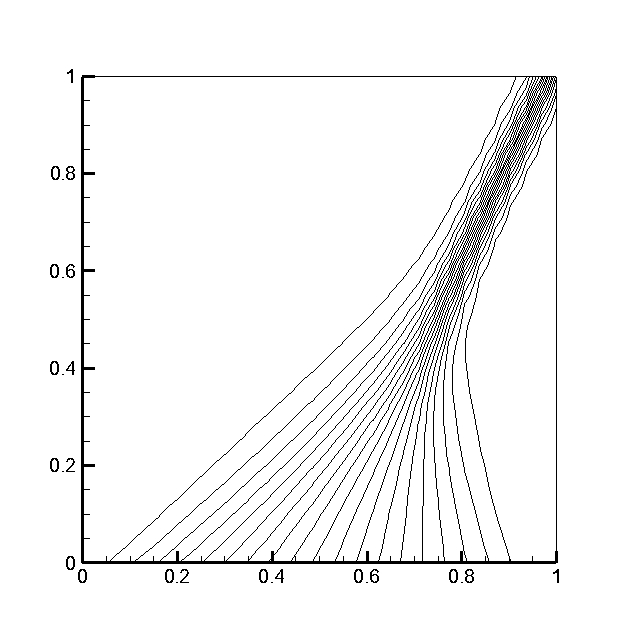}\includegraphics[scale=0.25]{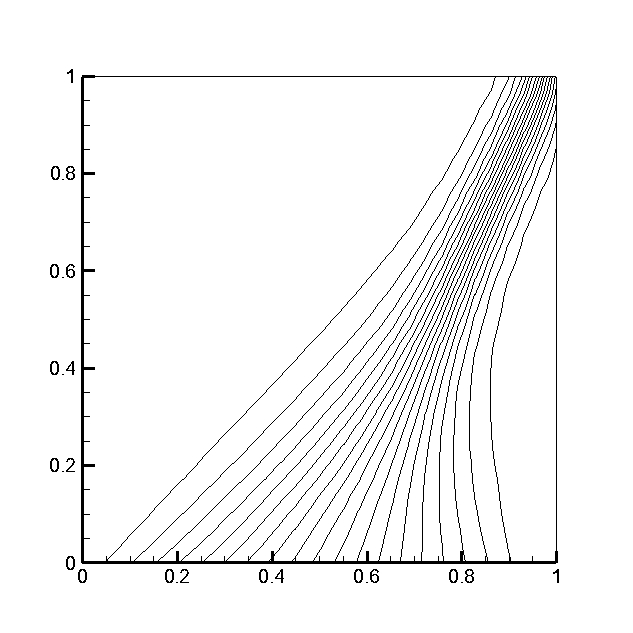}
\par\end{centering}

\begin{centering}
\includegraphics[scale=0.25]{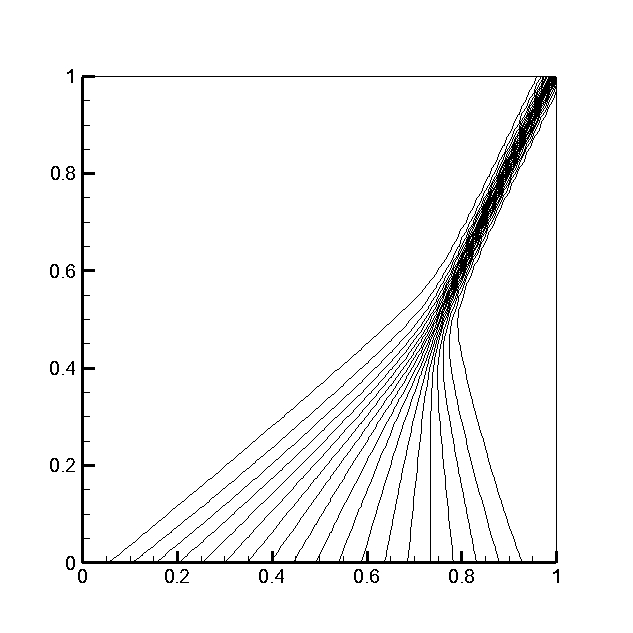}\includegraphics[scale=0.25]{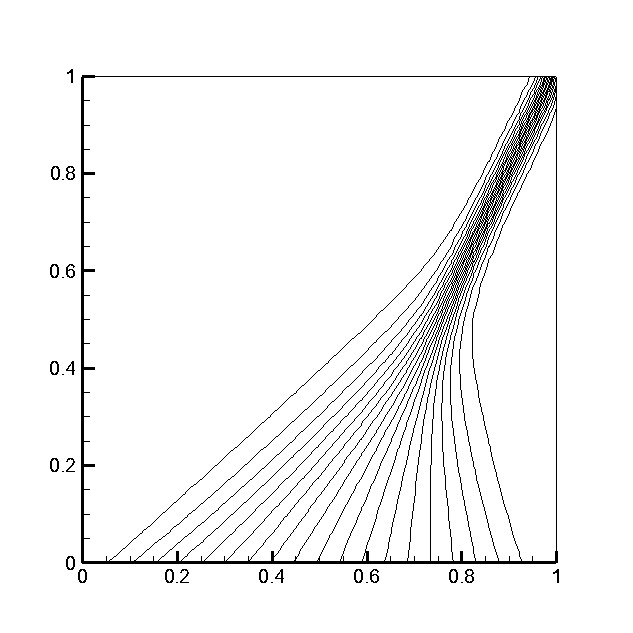}
\par\end{centering}

\caption{ Contours of solution to the oblique shock test case for 2D Burger's
equation by GINEUS (left) compared with KFVS (right) on 32$\times$32(top)
and 64$\times$64 (bottom) grid {[}I order{]}.\label{fig:-BU1}}
\end{figure}

\begin{figure}
\begin{centering}
\includegraphics[scale=0.25]{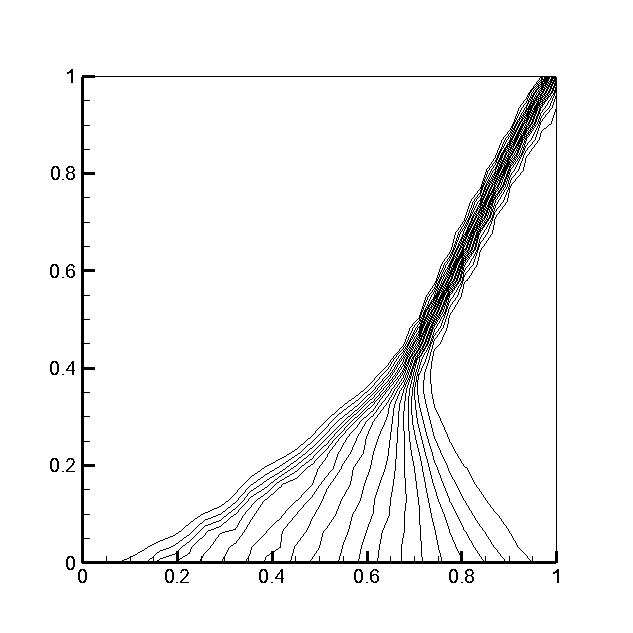}\includegraphics[scale=0.25]{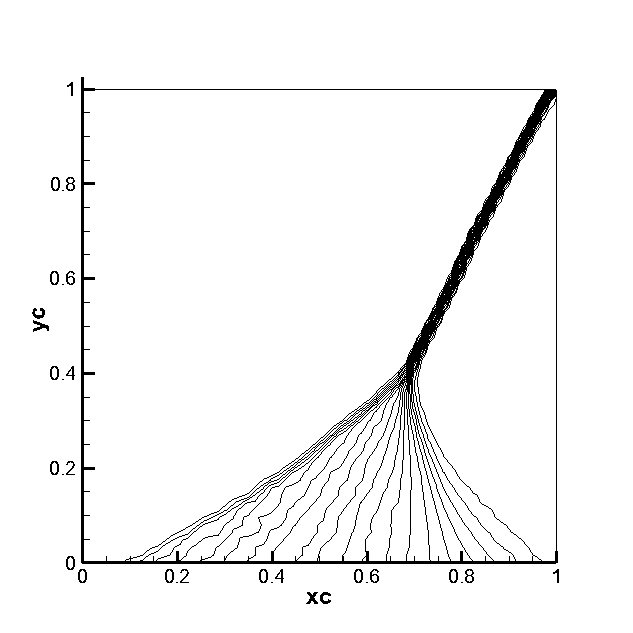}
\par\end{centering}

\caption{Contours of solution to the oblique shock test case for 2D Burger's
equation by GINEUS on 32$\times$32 (left) and 64$\times$64 (right)
grid {[}II order{]}.\label{fig:BU2}}
\end{figure}

First order results compared with KFVS(as shown in figure$(\ref{fig:-BU1})$)
shows genuinely multidimensional nature of the scheme as the shock
capturing is less diffusive even though shock is oblique to the grid
. Second order results are shown in figure $(\ref{fig:BU2})$.

\subsection{Regular Shock Reflection }

Specification:

Domain: (0,3)$\times$(0,1)

Grid : Uniform 60$\times$20, 120$\times$40

Initial Conditions: $(\rho,u,v,p)\mid_{(x,y,0)}$= (1, 2.9, 0, 1/1.4) 

Boundary Conditions:

$(\rho,u,v,p)\mid_{(0,y,t)}$= (1, 2.9, 0, 1/1.4) 

$(\rho,u,v,p)\mid_{(x,1,t)}$ = (1.69997, 2.61934, 0.50633, 1.52819) 

Dirichlet boundary condition is applied on top left boundaries. Bottom
boundary is perfectly reflecting wall and right boundary is supersonic
outlet.

\begin{figure}
\begin{centering}
\includegraphics[scale=0.4]{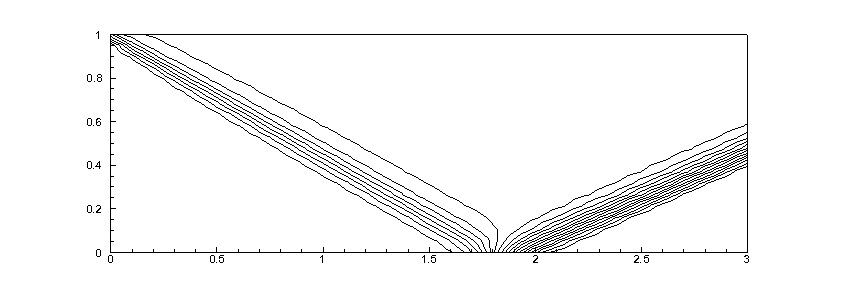}\includegraphics[scale=0.4]{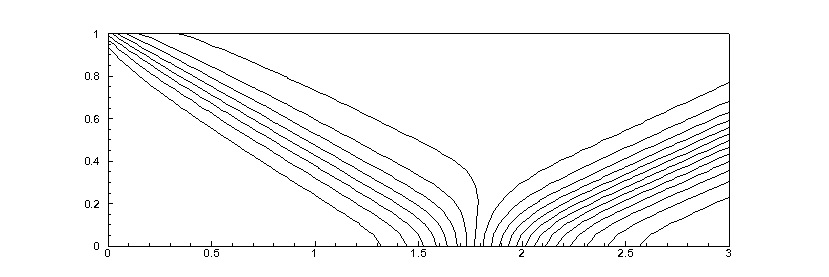}
\par\end{centering}

\begin{centering}
\includegraphics[scale=0.4]{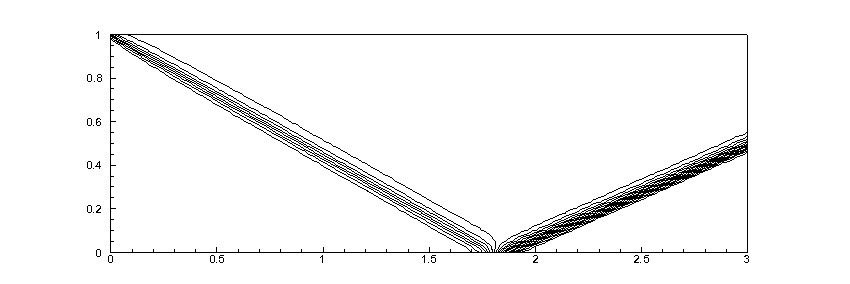}\includegraphics[scale=0.4]{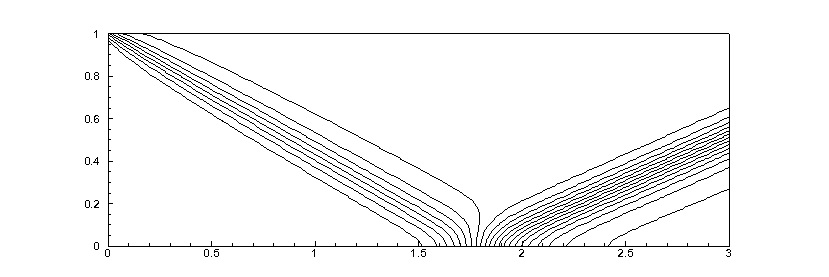}
\par\end{centering}

\centering{}\caption{Pressure contours by GINEUS (left) compared with KFVS (right) for
regular shock reflection in 60$\times$20 (top) and 120x40 (bottom)
grid {[}I order{]}.\label{fig:RSR1}.}
\end{figure}

\begin{figure}
\begin{centering}
\includegraphics[scale=0.4]{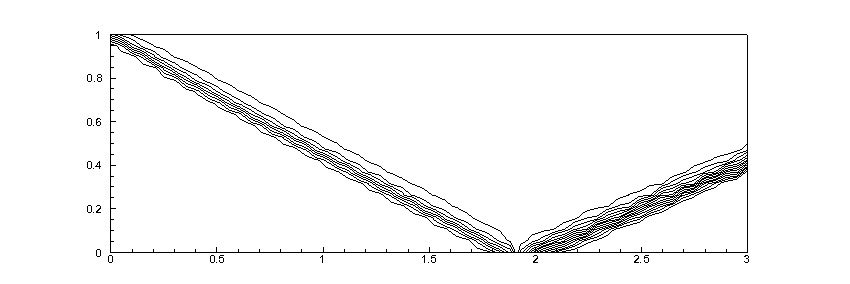}\includegraphics[scale=0.4]{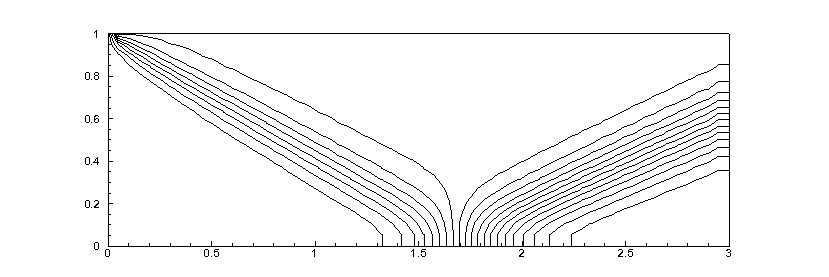}
\par\end{centering}

\begin{centering}
\includegraphics[scale=0.4]{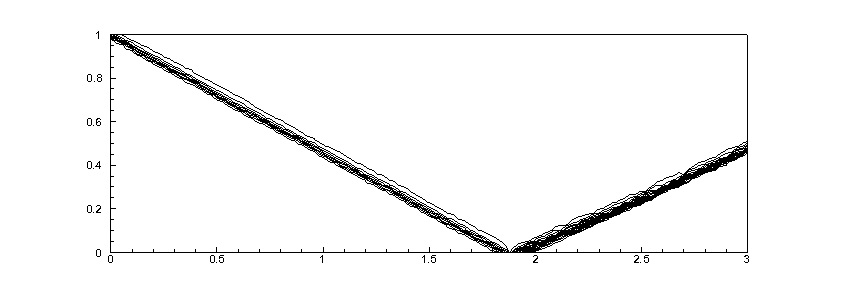}\includegraphics[scale=0.4]{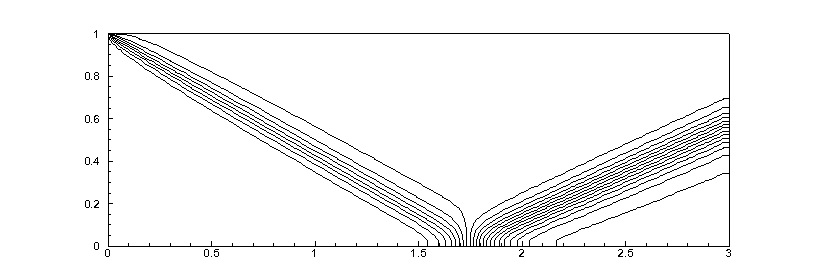}
\par\end{centering}

\caption{Pressure contours by GINEUS (left) compared with KFVS (right) for
regular shock reflection in 60$\times$20 (top) and 120x40 (bottom)
grid {[}II order{]}\label{fig:RSR2}.}
\end{figure}

The pressure values across the domain obtained with GINEUS are compared
with KFVS (as shown in figure $(\ref{fig:RSR1})$) and it becomes
clear that GINEUS has far better resolution of shock than KFVS. Similar
trend can be seen in the second order comparison (figure $(\ref{fig:RSR2})$).

\subsection{Horizontal Slip Flow}

Specification (\citet{Manna1992}):

Domain: (0,1)$\times$(0,1)

Grid : Uniform 40$\times$40

Initial Condition: M=2 for y<=x and M=3 for y>x.

Boundary Condition: Left boundary is supersonic inflow and top, bottom
and right boundary is supersonic outflow.

\begin{figure}
\begin{centering}
\includegraphics[scale=0.25]{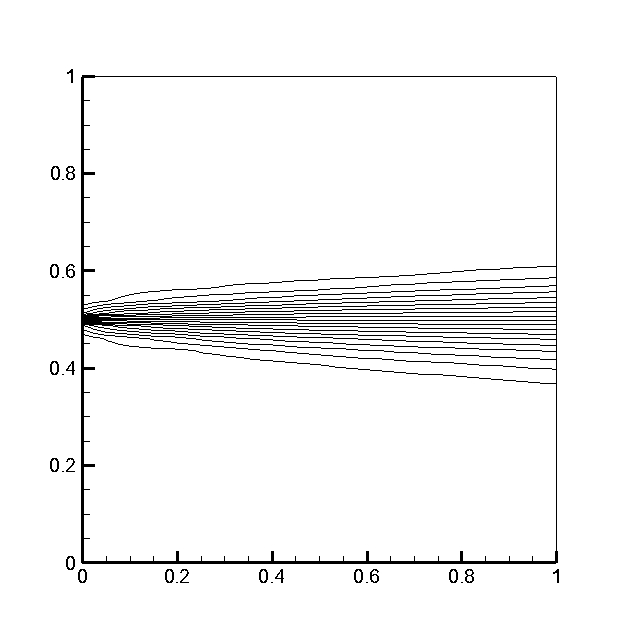}\includegraphics[scale=0.25]{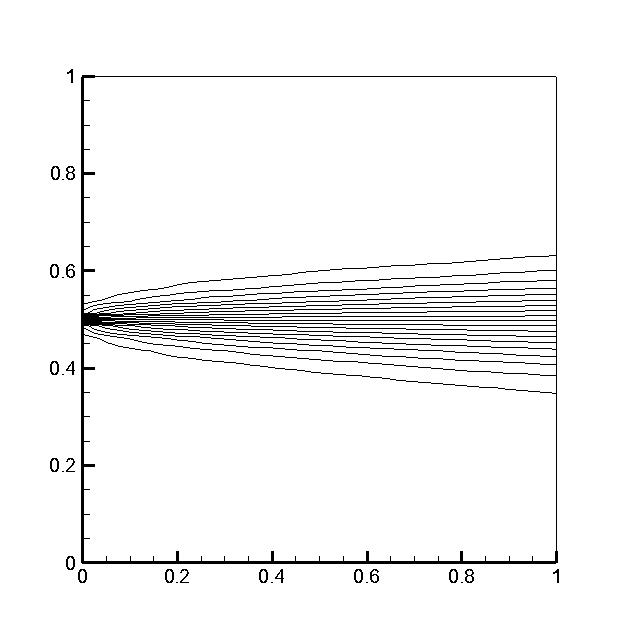}
\par\end{centering}

\caption{Mach contours for Horizontal Slip flow by GINEUS (left) and KFVS (right)
on 40$\times$40 grid.\label{fig:SF1}}
\end{figure}

To test the genuinely multidimensional nature of the scheme we are
comparing the resolution of contacts aligned to the grid and it is
found that there is no appreciable difference (figure $(\ref{fig:SF1})$).
In the next section oblique slip flow is compared and it is expected
that the resolution of contact with GINEUS must be same in both cases.

\subsection{Oblique Slip Flow}

Specification:

Domain: (0,1)$\times$(0,1)

Grid : Uniform 40$\times$40

Initial Condition: M=2 for y<=x and M=3 for y>x.

Boundary Condition: Left and bottom boundary is supersonic inflow
and top and right boundary is supersonic outflow.

\begin{figure}
\begin{centering}
\includegraphics[scale=0.34]{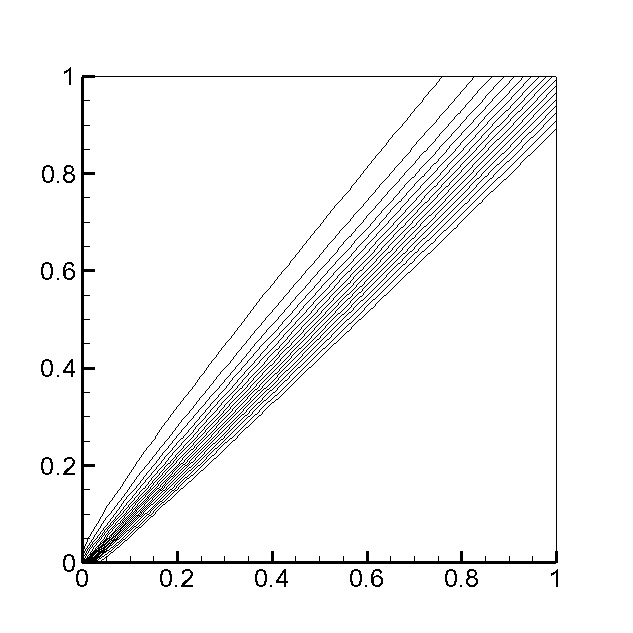}\includegraphics[scale=0.25]{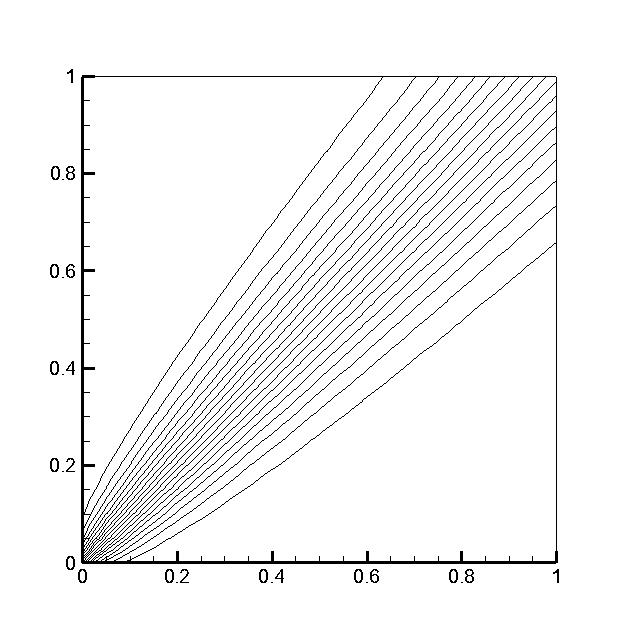}
\par\end{centering}

\begin{centering}
\includegraphics[scale=0.34]{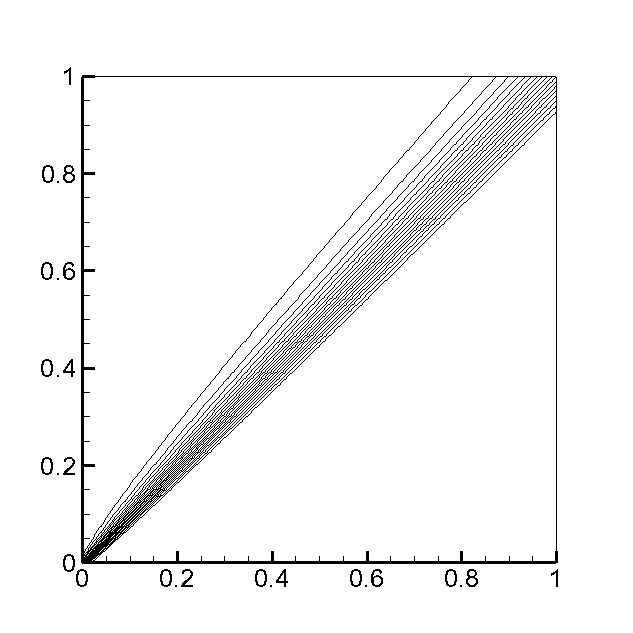}\includegraphics[scale=0.25]{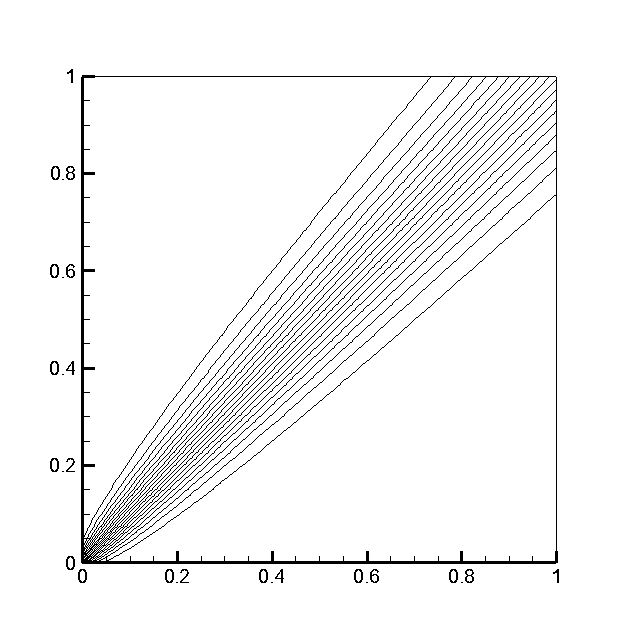}
\par\end{centering}

\caption{Mach contours by GINEUS (left) compared with KFVS (right) for oblique
slip flow in 40$\times$40 (top) and 80$\times$80 (bottom) grid {[}I
order{]}.\label{fig:OSF1}}
\end{figure}

It can be clearly seen that (figure $(\ref{fig:OSF1})$), for GINEUS,
resolution of shock is similar to the horizontal slip flow whereas
diffusion of contact has increased for KFVS.

\subsection{Supersonic Jet Interaction}

Specification (\citet{HDeconinck1993}):

Domain: (0,1)$\times$(0,1)

Grid : Uniform 40$\times$40

Initial Condition:

\begin{equation}
\begin{Bmatrix}M_{1}=4,\,\rho_{1}=0.5,\, p_{1}=0.25\,\, for\, y>0.5\\
M_{2}=2.4,\,\rho_{2}=1.0,\, p_{2}=1.0\,\, for\, y<0.5
\end{Bmatrix}
\end{equation}

Boundary Condition: Left boundary is supersonic inflow and top, bottom
and right boundary is supersonic outflow.

\begin{figure}
\begin{centering}
\includegraphics[scale=0.25]{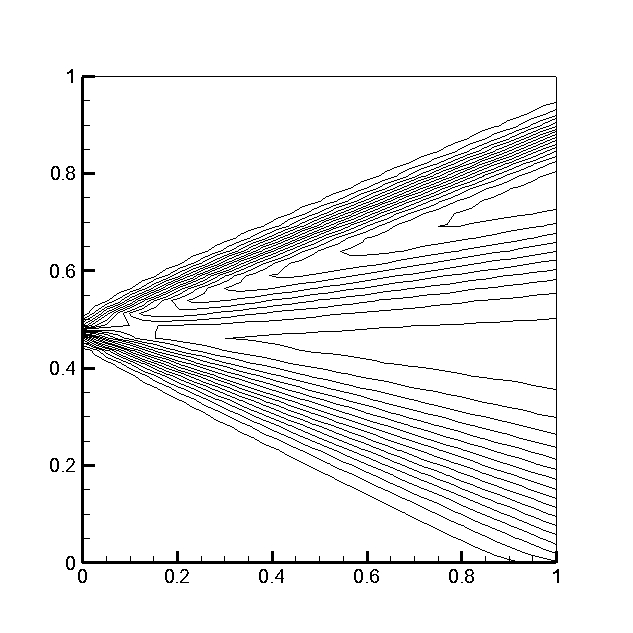}\includegraphics[scale=0.25]{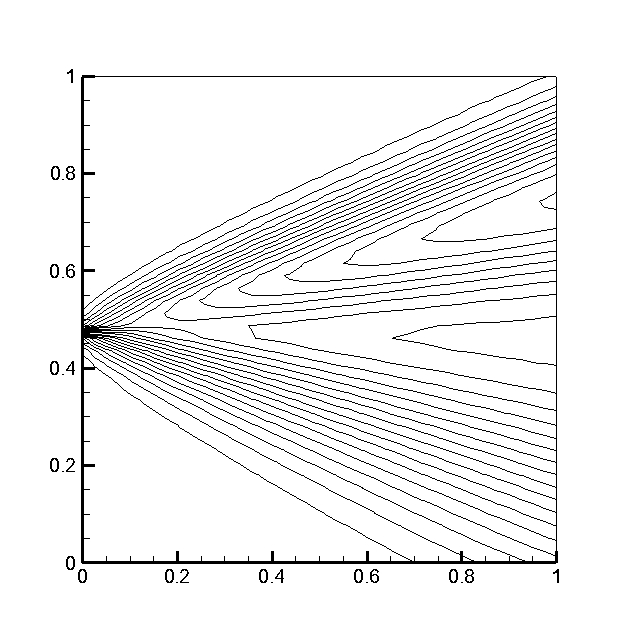}
\par\end{centering}

\begin{centering}
\includegraphics[scale=0.25]{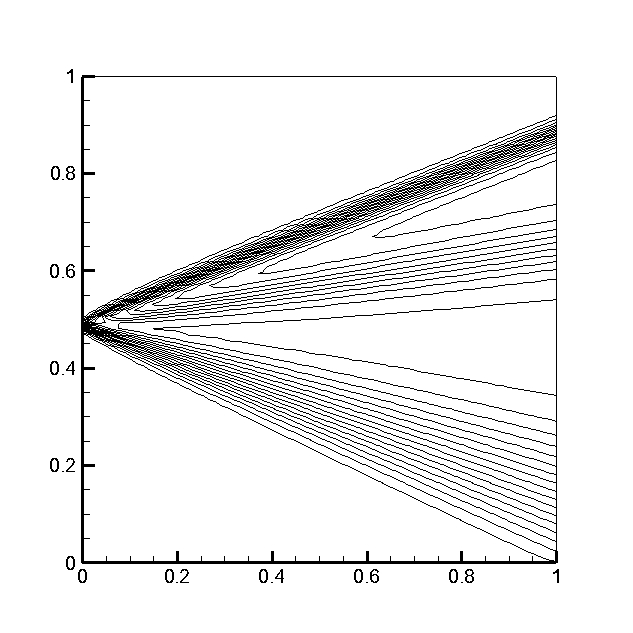}\includegraphics[scale=0.25]{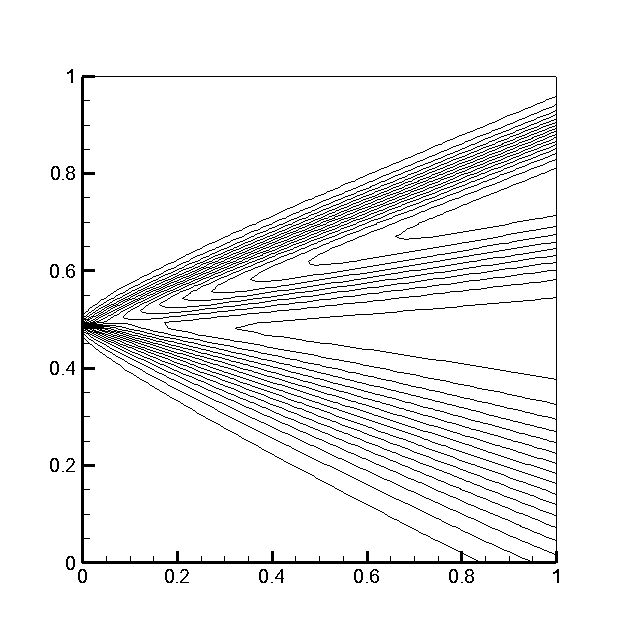}
\par\end{centering}

\caption{Density contours by GINEUS (left) compared with KFVS (right) for supersonic
jet reflection on 40$\times$40 (top) and 80$\times$80 (bottom) grid
{[}I order{]}.\label{fig:SJI1}}
\end{figure}

\begin{figure}
\begin{centering}
\includegraphics[scale=0.25]{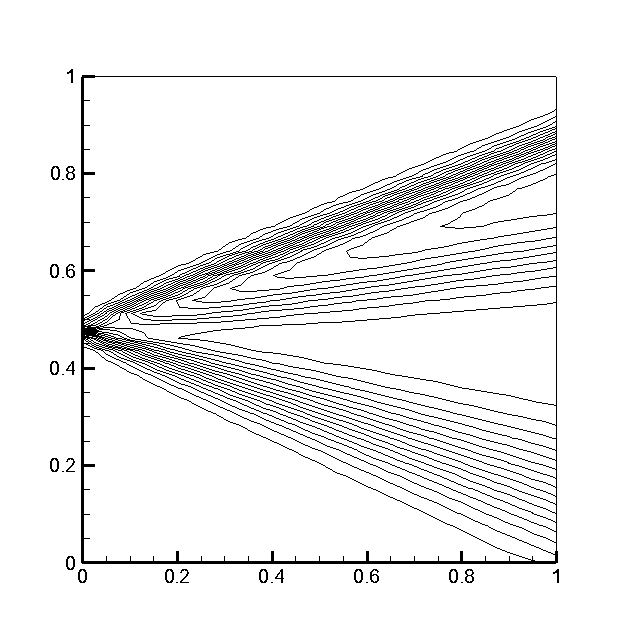}\includegraphics[scale=0.25]{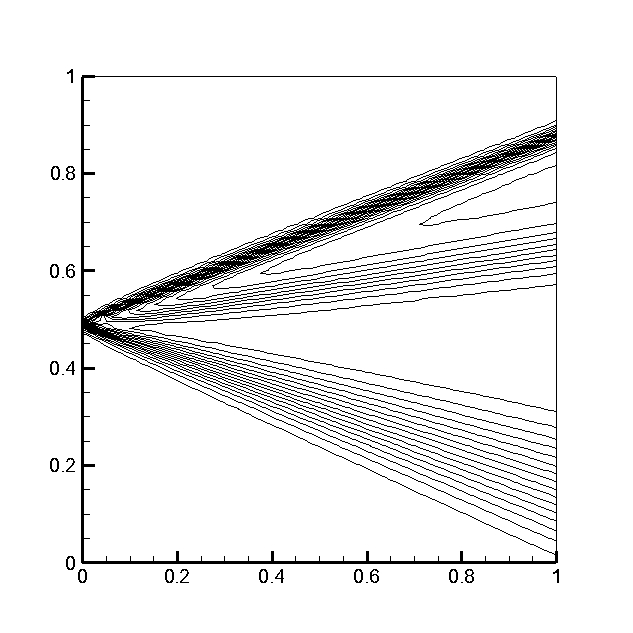}
\par\end{centering}

\caption{Density contours by GINEUS for supersonic jet reflection on 40$\times$40
(left) and 80$\times$80 (right) grid {[}II order{]}.\label{fig:SJI2}}
\end{figure}

The resolution of shock and expansion in case is much better in case
of GINEUS as compared with KFVS (figure $(\ref{fig:SJI1})$). Whereas
the resolution of contacts is nearly comparable because contacts are
nearly alligned to the grid. This is a perfect demonstration of multidimensional
feature of GINEUS. The density contours for second order extension
is given in figure $(\ref{fig:SJI2})$.

\subsection{Modified Supersonic Jet Interaction}

Specification:

Domain: (0,1)$\times$(0,1)

Grid : Uniform 40$\times$40

Initial Condition:

\begin{equation}
\begin{Bmatrix}M_{1}=4,\,\rho_{1}=0.5,\, p_{1}=0.08\,\, for\, y>0.5\\
M_{2}=2.8,\,\rho_{2}=1.0,\, p_{2}=1.3\,\, for\, y<0.5
\end{Bmatrix}
\end{equation}

Boundary Condition: Left boundary is supersonic inflow and top, bottom
and right boundary is supersonic outflow. 

\begin{figure}
\begin{centering}
\includegraphics[scale=0.3]{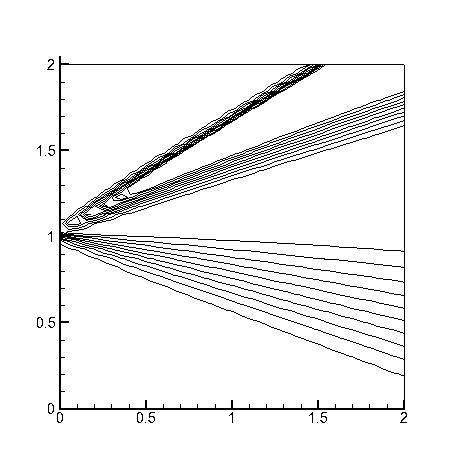}\includegraphics[scale=0.3]{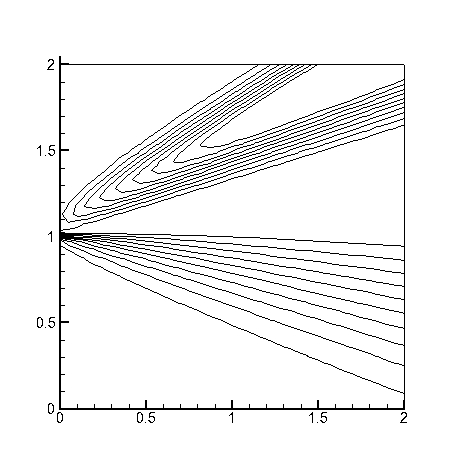}
\par\end{centering}

\caption{Density contours by GINEUS (left) compared with KFVS (right) for modified
supersonic jet reflection on 40$\times$40 grid\label{fig:MSF}.}

\end{figure}

This test case is created to demonstrate improved capturing of contacts
when it is not aligned to the grid.(figure $\ref{fig:MSF}$)

\subsection{Supersonic flow over a Ramp}

Specification (\citet{1993}):

Domain: (0,3)$\times$(0,1)

Grid : Uniform 120$\times$40. Since the scheme is meshless, we don't
need to switch to non-cartesian grid. Instead grid points are divided
into two categories: domain interior (flow region) and domain exterior.
The points in the domain exterior whose adjacent point is in domain
interior is taken as ghost points. The ramp starts at x=0.5 and extends
till x=1.0 inclined at an angle of $15^{o}$.

Initial Condition: Mach 2 flow inside the domain.

Boundary Condition: Left boundary is supersonic inflow. Top, bottom
and ramp surface is reflecting boundary. Right boundary is supersonic
outflow.

\begin{figure}
\begin{centering}
\includegraphics[scale=0.5]{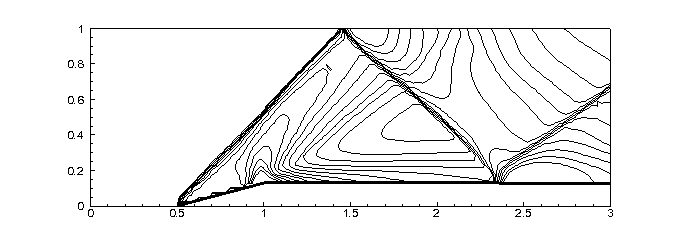}
\par\end{centering}

\caption{Mach Contours by GINEUS with I order accuracy for the flow over a
ramp test case on 120$\times$40 grid.\label{fig:R1}}
\end{figure}

Different features of the flow is brilliantly captured even with course
grid of 120$\times$40 (figure $(\ref{fig:R1})$).

\subsection{Shock Explosion in a Box}

Specification(\citeauthor{AMROC}):

Domian: (0,1)$\times$(0,1)

Grid : Uniform 50$\times$50 , 100$\times$100.

Initial Condition: Shock bubble of radius 0.3 centered at (0.4,0.4).
Pressure and density inside the bubble is 5 and outside it is 1. At
t=0, fluid is stationary. 

Boundary Condition: All boundaries are reflecting.

\begin{figure}
\begin{centering}
\includegraphics[scale=0.3]{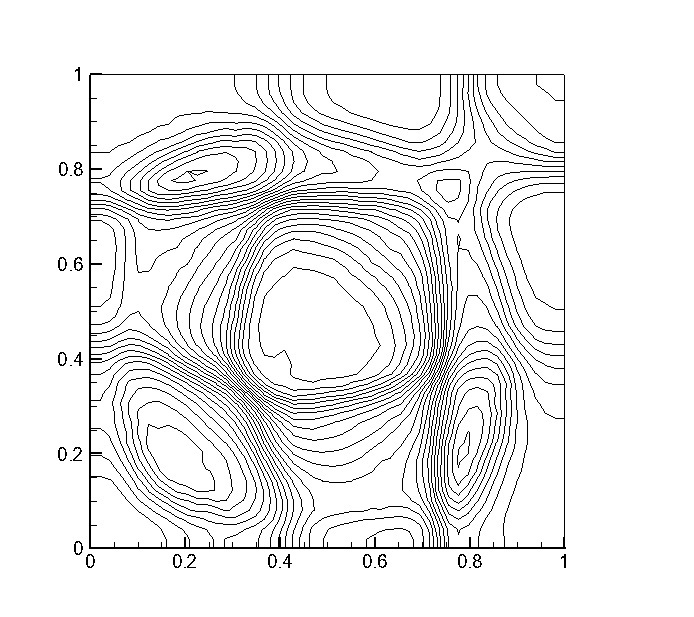}\includegraphics[scale=0.3]{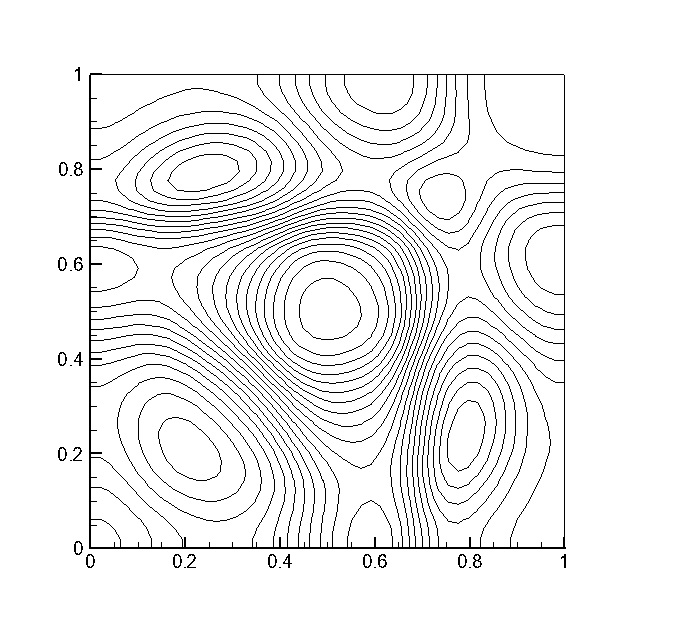}
\par\end{centering}

\begin{centering}
\includegraphics[scale=0.3]{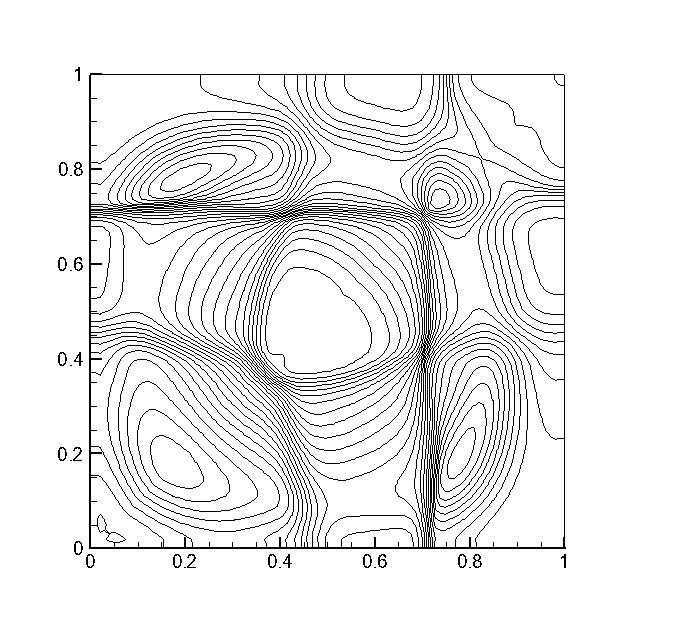}\includegraphics[scale=0.3]{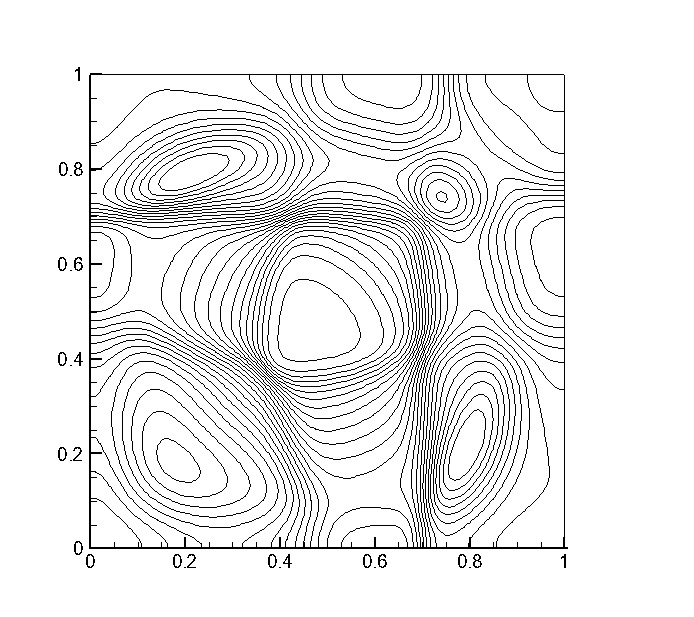}
\par\end{centering}

\caption{Density contours by GINEUS (left) compared with KFVS (right) for explosion
in a box on 50$\times$50 (top) and 100$\times$100 (bottom) grid
{[}I order{]}.\label{fig:EP1}}
\end{figure}

\begin{figure}
\begin{centering}
\includegraphics[scale=0.3]{EP_2_G_CFL_0p4_50X50_Time_0p5}\includegraphics[scale=0.3]{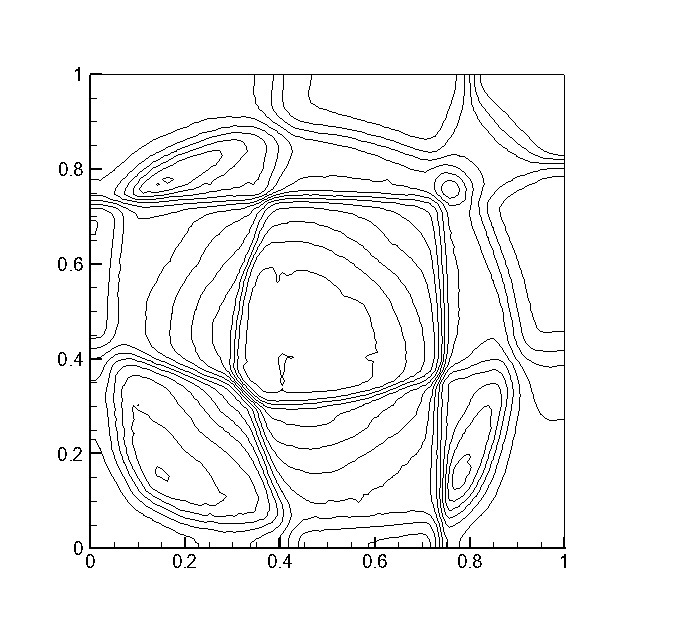}
\par\end{centering}

\caption{Density contours by GINEUS for explosion in a box on 50$\times$50
(left) and 100$\times$100 (right) grid {[}II order{]}.\label{fig:EP2}}
\end{figure}

The initial shock fronts will expand to the boundaries and get reflected.
These reflected waves will interact with each other to produce a complicated
flow field. The flow is observed after t=0.5s and density contours
for first order compared with KFVS can be seen in figure $(\ref{fig:EP1})$
and for second order in figure $(\ref{fig:EP2})$.

\subsection{Double Mach Reflection}

Specification (\citet{Woodward1984}) :

Domain: (0,4)$\times$(0,1)

Grid: 240$\times$60

Initial Condition: A shock of M=10 is present at an angle of $30^{0}$
to the x-axis, starting from x-axis at x=1/6. Density and pressure
ahead of the shock is 1.4 and 1 respectively.

Boundary Condition: The intersection of top boundary with shock front
moves with a speed of 10/cos($30^{0}$). Top boundary values ahead
and behind the shock assumes initial values. Left boundary is supersonic
inflow wheras right boundary is simply extrapolation from inside.
The bottom boundary before x=1/6 is constant initial value and for
x>1/6, it is a reflecting wall.

\begin{figure}
\begin{centering}
\includegraphics[scale=0.4]{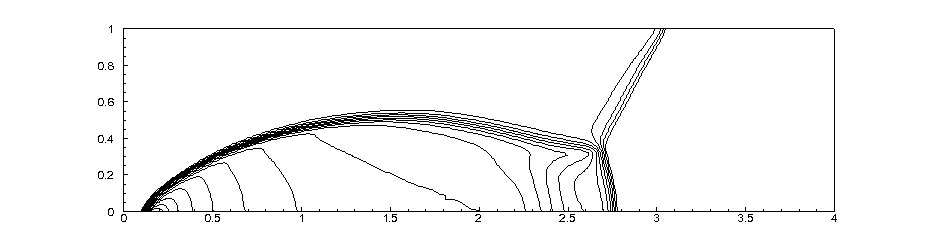}
\par\end{centering}

\begin{centering}
\includegraphics[scale=0.4]{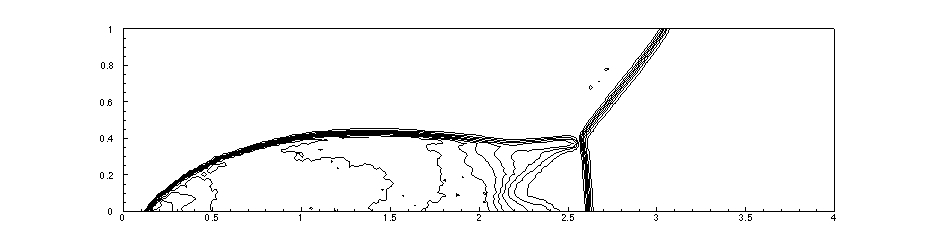}
\par\end{centering}

\caption{Pressure Contours by GINEUS having I order (top) and II order (bottom)
accuracy for Double Mach Reflection on 240$\times$60 grid\label{fig:DMR12}}
\end{figure}

The flow features are observed after t=0.2s. Pressure contours for
the first and second order GINEUS is plotted in figure ($\ref{fig:DMR12}$).

\subsection{2D Riemann Problem}

Specification (\citet{Liska}):

Domain: (0,1)$\times$(0,1)

Grid: 100$\times$100, 200$\times$200

Initial Condition: Jump in primitive variables in 2D

\begin{equation}
\begin{cases}
\rho=0.5065,\, u_{1}=0.8939,\, u_{2}=0.0,\, p=0.35 & \, for\, x<0.5\, y>0.5\\
\rho=1.1,\, u_{1}=0.0,\, u_{2}=0.0,\, p=1.1 & \, for\, x\geq0.5\, y>0.5\\
\rho=1.1,\, u_{1}=0.8939,\, u_{2}=0.8939,\, p=1.1 & \, for\, x<0.5\, y\leq0.5\\
\rho=0.5065,\, u_{1}=0.0,\, u_{2}=0.8939,\, p=0.35 & \, for\, x\geq0.5\, y\leq0.5
\end{cases}
\end{equation}

Boundary Condition: All the boundary values are obtained simply by
extrapolating from inside. 

The flow is observed after t=0.25s.

\begin{figure}
\begin{centering}
\includegraphics[scale=0.3]{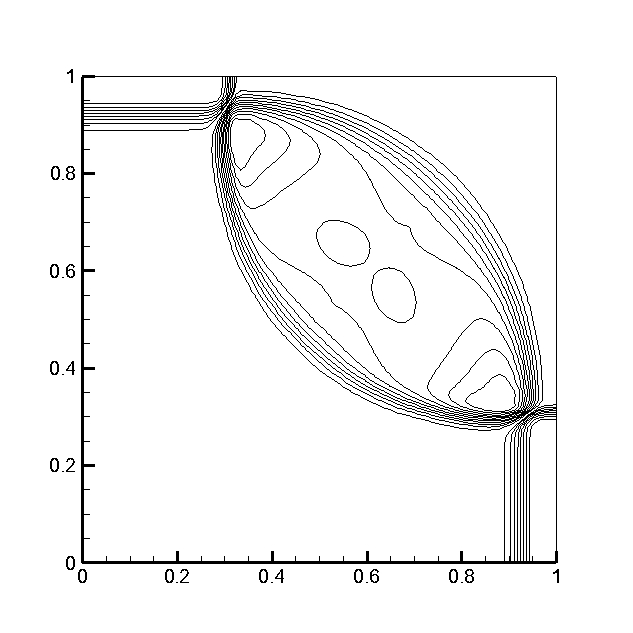}\includegraphics[scale=0.3]{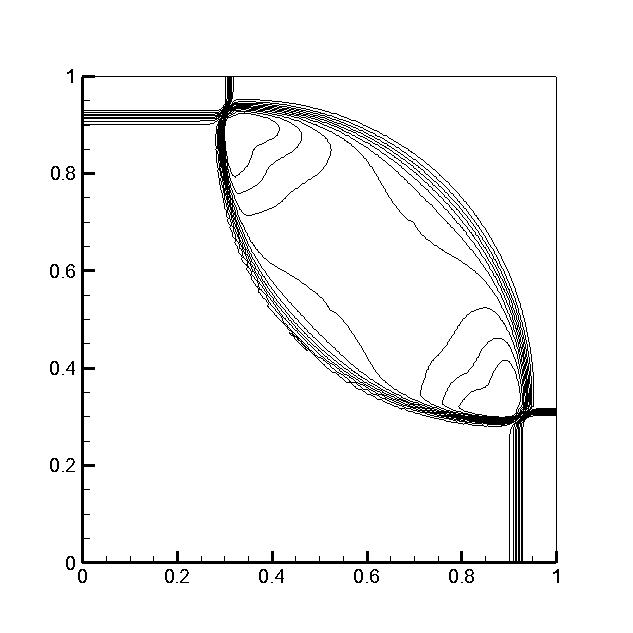}
\par\end{centering}

\caption{Density contours of GINEUS with I order accuracy for 2D Riemann test
case on 100$\times$100 (left) and 200$\times$200 (right) grid \label{fig:RP1}}
\end{figure}

\subsection{Some Additional 1D Riemann Problem}

\begin{figure}
\begin{centering}
\includegraphics[scale=0.5]{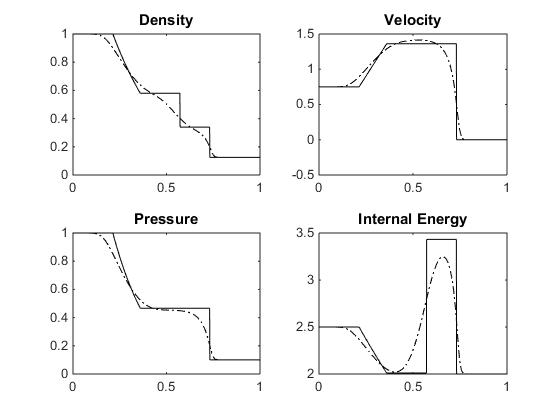}
\par\end{centering}

\caption{Torro Test case 1.\label{fig:TC1}}
\end{figure}

\begin{figure}
\begin{centering}
\includegraphics[scale=0.5]{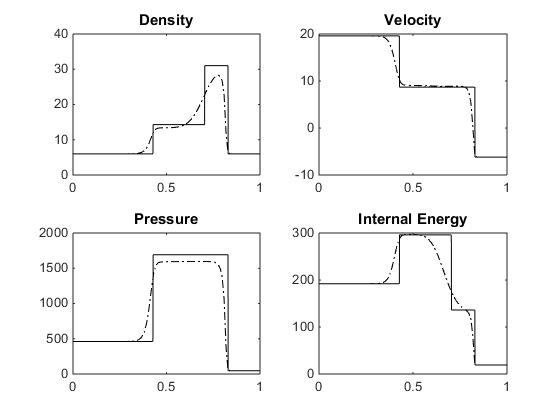}
\par\end{centering}

\caption{Torro Test case 4.\label{fig:TC4}}
\end{figure}

The scheme is readily built to capture flow feautures in multidimensions
but to show its robustness, this 2D formulation is used to model a
1D flow. This is done by making the domain in y-direction very small
while modelling the flow in x-direction and using symmetric boundary
condition for top and bottom walls. The 1D test case used is Torro's
Riemann Problem (\citet{Toro2009}) and primitive variables are plotted
for test case 1 and test case 4 (figure $(\ref{fig:TC1})$ and figure
$(\ref{fig:TC4})$ respectively) . 

\section*{5 Conclusion}

As mentioned in the first section, present formulation is implementation
of the framework with upwind method. The framework can be extended
to any other finite volume, finite difference or finite element scheme.
In order to obtain extra data points for the second order, we can
replace 'Arc of Approach' method by second order polynomial approximation
of the distribution function inside a cell, i.e.

\begin{equation}
f(x,y)=a_{1}x^{2}+a_{2}y^{2}+a_{3}xy+a_{4}x+a_{5}y+a_{6}.
\end{equation}

Suppose that there are $'n'$ data points around the central data
point. Determining the neighboring points in a mesh-less domain is
itself tricky but there are established methods to resolve this issue.
To calculate the coefficient $a_{i}s$, we minimize the squared error
given by $\ref{eq:Sqerror}$, with respect to each of $a_{i}$. 

\begin{equation}
S=\sum_{i=0}^{n}(f(x_{i},y_{i})-f_{i}^{eq})^{2}\label{eq:Sqerror}
\end{equation}

Using this interpolated function, we can get the data value inside
the region confined by the neighbouring points which can be used in
the second order extension of the scheme.

Apart from the variations in the discretization, distribution function
can also be varied. Perthame's hat function\citep{Perthame1991} or
Qu, Shu and Chew's circular distribution function \citep{KunQu2007}
are the possible replacements for the beams.\pagebreak{}

\bibliographystyle{apalike}
\bibliography{references}

\pagebreak{}

\appendix

\section{Linear Interpolation of Distribution Function}

The expression for $f_{q}$ in terms of $f_{r}\, and\, f_{s}$ can
be determined using linear interpolation as:

\begin{equation}
f_{q}=k_{1}r+k_{2}
\end{equation}

Using, $r$ = 0 at $f_{q}=f_{S}$ and $r$ = SR at $f_{q}=f_{R}$,
we have

\begin{equation}
f_{q}=\frac{(f_{R}-f_{S})}{SR}r+f_{S}\label{eq:Linear_Int}
\end{equation}

Now, to determine $r$, position of $P'$(point of intersection of
perpendicular drawn from the center to the line $RS$) has to be taken
into account. They are

\subsection{When $P'$ lie on $RS$:}

\begin{figure}
\begin{centering}
\includegraphics[scale=0.2]{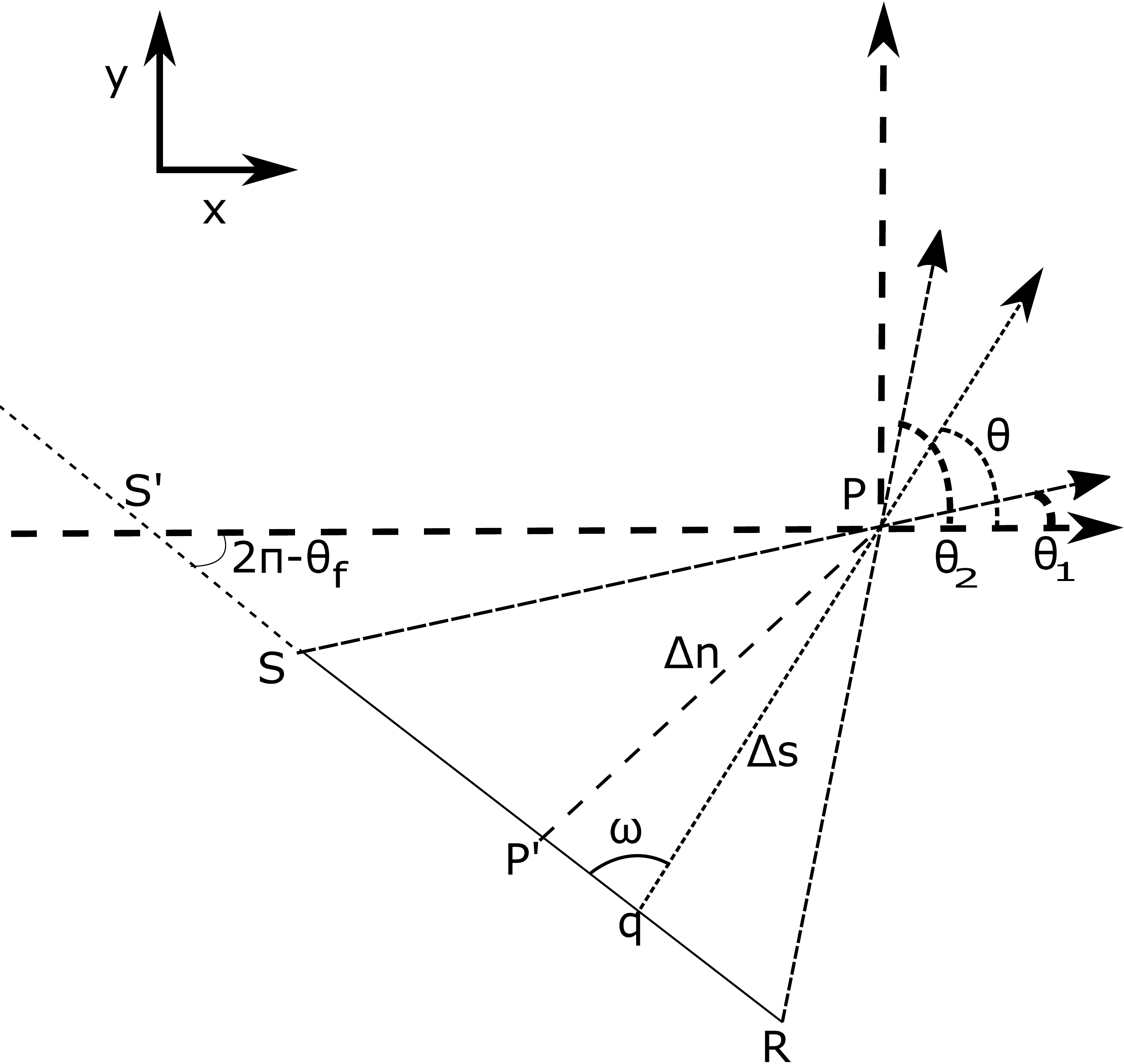}
\par\end{centering}

\caption{Parameter Definitions when P' lies in between RS.\label{fig:BRS}}
\end{figure}

Using the figure ($\ref{fig:BRS}$) :

\begin{equation}
\angle PS'S+\angle S'PS=\angle PqP'
\end{equation}

or $(2\pi-\theta_{f})+\theta=\omega$, Hence

\begin{equation}
\omega=\theta-\theta_{f}
\end{equation}

`

Now, 

\begin{equation}
r=Sq=SP'-qP'
\end{equation}

But $qP'=\bigtriangleup n/tan\omega$. Using $(\ref{eq:Linear_Int})$,

\begin{equation}
f_{q}=\frac{(f_{R}-f_{S})}{SR}(SP'-\frac{\bigtriangleup n}{tan\omega})+f_{S}
\end{equation}

which gives,

\begin{equation}
f_{q}=f_{S}(1-\frac{SP'}{SR}+\frac{\bigtriangleup n}{SR\times tan\omega})+f_{R}(\frac{SP'}{SR}-\frac{\bigtriangleup n}{SR\times tan\omega})
\end{equation}

so that

\begin{equation}
D_{1}=\frac{SP'}{SR}-\frac{\bigtriangleup n}{SR\times tan\omega}\,\,;\,\, D_{2}=1-\frac{SP'}{SR}+\frac{\bigtriangleup n}{SR\times tan\omega}
\end{equation}

\textbf{ }

\subsection{When $P'$ lie beyond $R$:}

\begin{figure}
\begin{centering}
\includegraphics[scale=0.2]{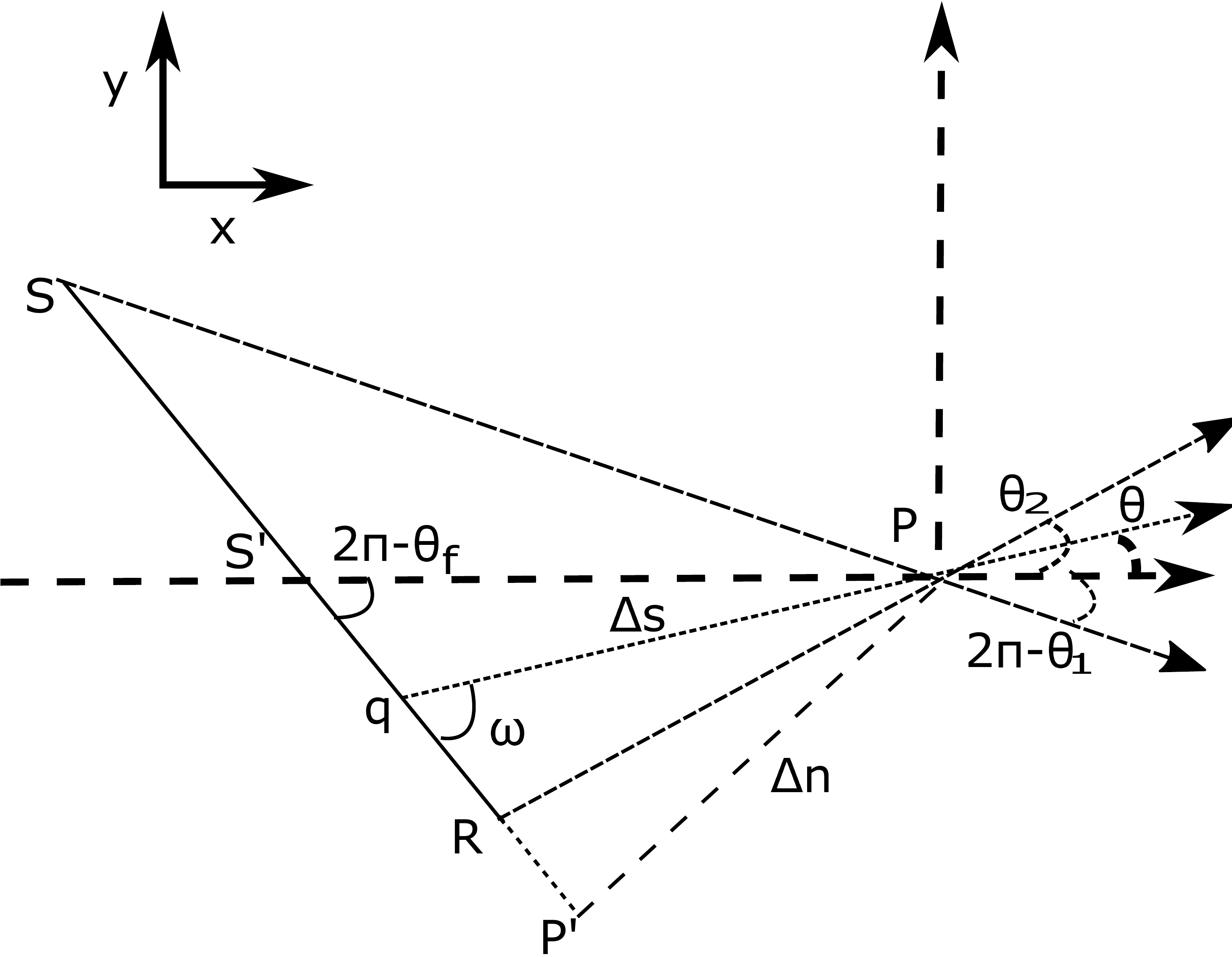}
\par\end{centering}

\caption{Parameter Definitions when P' lies beyond R.\label{fig:BR}}
\end{figure}

Using the figure ($\ref{fig:BR}$) :
\begin{equation}
\angle PS'q+\angle S'Pq=\angle PqR
\end{equation}

or $(2\pi-\theta_{f})+\theta=\omega$. Hence

\begin{equation}
\omega=\theta-\theta_{f}
\end{equation}

Now,

\begin{equation}
r=Sq=SP'-qP'.
\end{equation}

But $qP'=\bigtriangleup n/tan\omega$, hence

\begin{equation}
f_{q}=\frac{(f_{R}-f_{S})}{SR}(SP'-\frac{\bigtriangleup n}{tan\omega})+f_{S},
\end{equation}

which gives

\begin{equation}
f_{q}=f_{S}(1-\frac{SP'}{SR}+\frac{\bigtriangleup n}{SR\times tan\omega})+f_{R}(\frac{SP'}{SR}-\frac{\bigtriangleup n}{SR\times tan\omega}),
\end{equation}

so that

\begin{equation}
D_{1}=\frac{SP'}{SR}-\frac{\bigtriangleup n}{SR\times tan\omega}\,\,;\,\, D_{2}=1-\frac{SP'}{SR}+\frac{\bigtriangleup n}{SR\times tan\omega}.
\end{equation}

\subsection{When $P'$ lie beyond $S$:}

\begin{figure}
\begin{centering}
\includegraphics[scale=0.2]{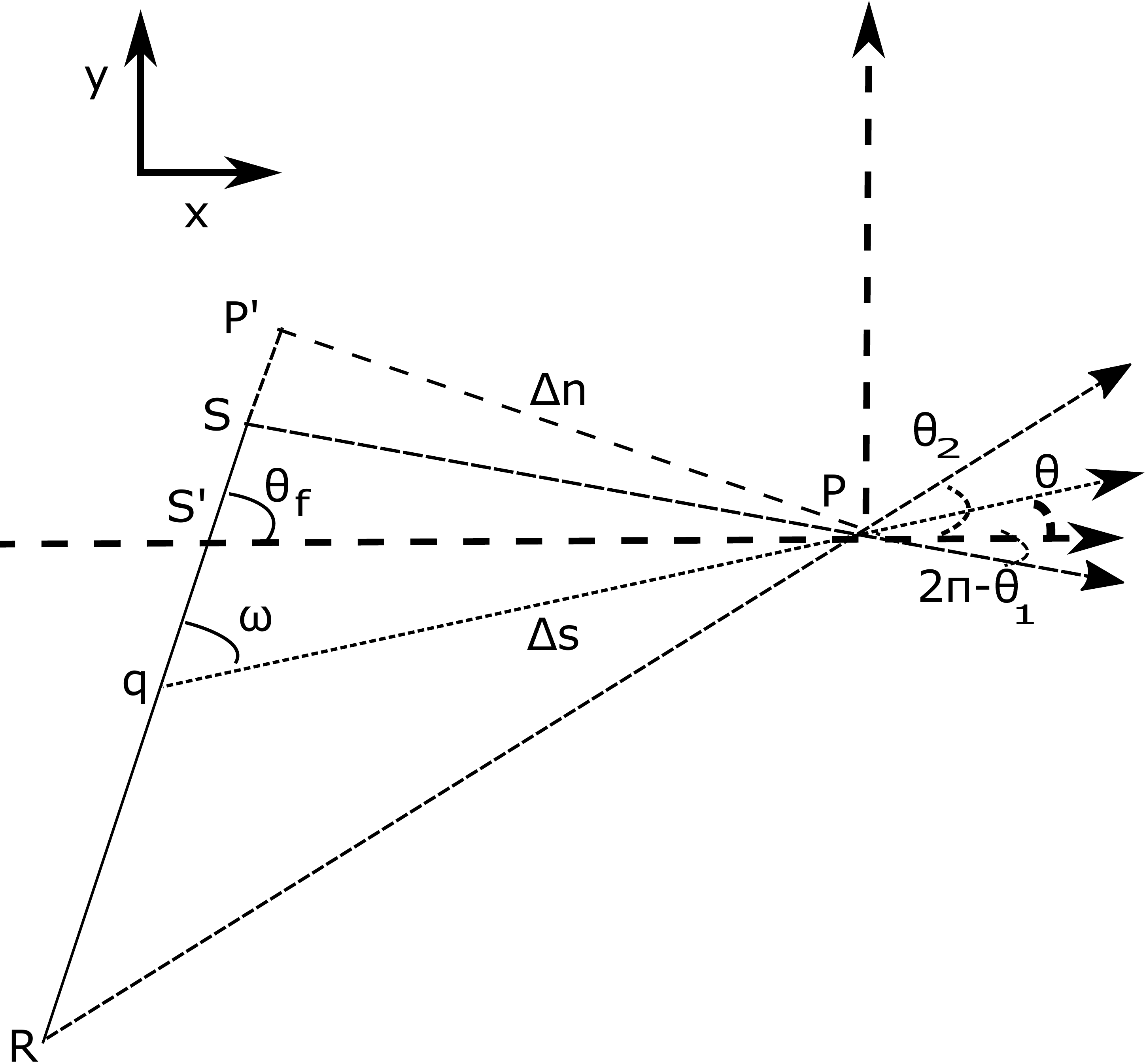}
\par\end{centering}

\caption{Parameter Definitions when P' lies beyond S\label{fig:BS}.}
\end{figure}

Using the figure ($\ref{fig:BS}$) :

\begin{equation}
\angle PS'q+\angle S'Pq+\angle S'qP=\pi
\end{equation}

or $(\pi-\theta_{f})+\theta+\omega=\pi$, hence

\begin{equation}
\omega=\theta_{f}-\theta.
\end{equation}

Now, 

\begin{equation}
r=Sq=qP'-SP'
\end{equation}

But, $qP'=\bigtriangleup n/tan\alpha$, hence

\begin{equation}
f_{q}=\frac{(f_{R}-f_{S})}{SR}(\frac{\bigtriangleup n}{tan\alpha}-SP')+f_{S}
\end{equation}

which gives

\begin{equation}
f_{q}=f_{S}(1+\frac{SP'}{SR}-\frac{\bigtriangleup n}{SR\times tan\omega})+f_{R}(\frac{\bigtriangleup n}{SR\times tan\omega}-\frac{SP'}{SR}),
\end{equation}

so that

\begin{equation}
D_{1}=\frac{\bigtriangleup n}{SR\times tan\omega}-\frac{SP'}{SR}\,\,;\,\, D_{2}=1+\frac{SP'}{SR}-\frac{\bigtriangleup n}{SR\times tan\omega}.
\end{equation}

\section{Replacing 2-D Maxwellian with Dirac-Delta Function}

We will be comparing and matching different moments of the Maxwellian
and Dirac Delta Function

The values for parameters appearing in the modified distribution function
can be obtained using moment matching, i.e.,

\begin{equation}
<f^{M}>=\rho\Rightarrow<f^{B}>=a+4b=\rho
\end{equation}

Similarly,

\begin{equation}
<v_{1}^{2}f^{B}>=au_{1}^{2}+4bu_{1}^{2}+2b\bigtriangleup u_{1}^{2}=P+\rho u_{1}^{2},
\end{equation}

\begin{equation}
<v_{2}^{2}f^{B}>=au_{2}^{2}+4bu_{2}^{2}+2b\bigtriangleup u_{2}^{2}=P+\rho u_{2}^{2},
\end{equation}

and

\begin{equation}
<(v_{1}-u_{1})^{4}f^{B}>=2b\bigtriangleup u_{1}^{4}=\frac{3P^{2}}{\rho}
\end{equation}

Solving the above four equations we get, $a$ =$\rho/3$, $b$ = $\rho/6$
and $\bigtriangleup u_{1}=\bigtriangleup u_{2}=\sqrt{3P/\rho}$.
\end{document}